**Title:** Controlled non-volatile modulation of optical dispersion in monolayer tungsten disulfide via ferroelectric polarization patterning

**Authors:** Yuhong Cao[1], Zekun Hu[1], Jason Lynch[1], Bongjun Choi[1], Kyung Min Yang[1], Hyunmin Cho[1], Chloe Leblanc[1], Chen Chen[2], Joan M. Redwing[2], Deep Jariwala[1]*

[1]Department of Electrical and Systems Engineering, University of Pennsylvania, Philadelphia, Pennsylvania 19104, USA

[2]2D Crystal Consortium Materials Innovation Platform, Materials Research Institute, Penn State University, University Park, PA, USA

*Corresponding author: Deep Jariwala (email: dmj@seas.upenn.edu)

**Abstract:**

The manipulation of optical properties, including reflection, refraction, polarization, phase, and frequency, has long been central to advancing photonic and optoelectronic technologies. However, existing electro-optical approaches rely on volatile mechanisms that require continuous power consumption. Here, we demonstrate strong, nonvolatile modulation of optical dispersion in monolayer tungsten disulfide (ML $WS_2$) using patterned ferroelectric domains in aluminum scandium nitride (AlScN). By locally poling ferroelectric domains into opposite states, we achieve substantial manipulation of the complex refractive index ($\Delta n > 0.7$, $\Delta k > 0.4$) and excitonic energy shifts (~50 meV) in ML $WS_2$, comparable to previous gate-tuning approaches while eliminating continuous power consumption. We introduce an asymmetric screening model that reveals how ferroelectric polarization induces carrier-density-dependent Coulomb screening, leading to distinct excitonic behaviors between electron- and hole-doped regions. Furthermore, we demonstrate a gate-free lateral p-n homojunction with a rectification ratio of $6\times10^3$, formed through spatial carrier redistribution. These findings establish ferroelectric/2D heterostructures as a powerful platform for nonvolatile optical dispersion engineering, enabling energy-efficient, reconfigurable photonic and optoelectronic devices.

**Introduction:**

Two-dimensional transition metal dichalcogenides (TMDCs) exhibit exceptional excitonic properties that can be electrically tuned for photonic applications[1-3]. Monolayer $WS_2$, with its direct bandgap in the visible spectrum and large exciton binding energy (~0.32 eV), presents an ideal platform for exploring light-matter interactions[4-7].

Extensive efforts have focused on modulating excitonic behavior in TMDCs through external stimuli, including gate voltage[8-10], strain[11-13], magnetic fields[14], optical control[15], and chemical doping[16]. However, these approaches suffer from fundamental limitations: gate tuning requires continuous power consumption and voltages exceeding 100 V, while chemical doping is typically irreversible. These constraints severely limit practical device implementation.

Ferroelectric materials offer a promising solution through nonvolatile, switchable polarization states. Previous ferroelectric gating demonstrations in TMDCs have shown programmable junctions[17-19] and nanoscale exciton control[20-23]. However, these studies have primarily relied on photoluminescence observations, leaving the fundamental opportunity for nonvolatile optical dispersion modulation unexplored. Furthermore, the physical mechanisms governing ferroelectric-tuned excitonic behavior remain poorly understood.

Aluminum scandium nitride (AlScN) has recently emerged as an exceptional ferroelectric material with remnant polarization exceeding 100 µC/cm²—significantly surpassing conventional ferroelectrics[24-26]. This unprecedented polarization strength, combined with CMOS compatibility, positions AlScN as an ideal platform for exploring ferroelectric control of 2D materials.

In this work, we demonstrate strong non-volatile tuning of the full optical dispersion in ML $WS_2$ by employing patterned AlScN ferroelectric domains. Using imaging spectroscopic ellipsometry (ISE), we observe significant modulation of the complex refractive index and excitonic peak shift of ML $WS_2$ under different ferroelectric polarization states. An asymmetric screening model is first proposed to elucidate the ferroelectric-tuned excitonic behavior. Additionally, the system exhibits the formation of a lateral p-n homojunction via gate-free electrostatic doping, further revealing the unique capabilities of ferroelectric domain engineering. Our work establishes a promising experimental platform for nonvolatile modulation of optical constants and carrier distribution, opening avenues for dynamic light manipulation in future nonvolatile photonic and optoelectronic applications.

**Results:**

**Characterization of functionalized AlScN patterns and ML $WS_2$/AlScN device**

The initial AlScN ferroelectric dielectric film used in this study is deposited on an Al(111)/Sapphire ($Al_2O_3$) substrate, with a 50 nm aluminum (Al) top layer, as illustrated in Fig. S1a. The original top Al layer is subsequently etched and replaced with newly deposited, spatially separated Al electrodes (Figs. S1b and S1c) to define regions with different polarization states (Fig. S1d). Figure 1a schematically illustrates the electrical poling process of the ferroelectric AlScN film. By applying triangular voltage pulses of opposite polarity to the separate Al electrodes, the AlScN domains are locally polarized either upward ($P_{up}$) or downward ($P_{down}$) (see Supplementary Methods and Fig. S2 for details). After poling, the Al electrodes are removed again via etching (Fig. S1e), and a ML $WS_2$ flake prepared via mechanical exfoliation is transferred onto the poled AlScN pattern (Supplementary Methods and Fig. S1f). Figure 1b presents the schematic patterned AlScN, and Fig. 1c shows the structure of the ML $WS_2$/patterned AlScN device.

Figure 1g displays an optical micrograph of the fabricated ML $WS_2$/AlScN device. The inset shows the three individually patterned Al electrodes on the AlScN surface. The triangular, semi-transparent flake represents the ML $WS_2$ transferred across the polarized AlScN regions. Regions 1, 2, and 3 correspond to $P_{up}$, $P_{down}$, and $P_{up}$, respectively. Yellow markers are added to aid visual identification. To confirm that the AlScN is well-poled, piezoresponse force microscopy (PFM) is employed to image the phase contrast across the surface (Fig. 1d), revealing a clear 180° phase shift between oppositely polarized regions[27], as shown in the bottom curve. Additionally, scanning Kelvin probe microscopy (SKPM) is utilized to directly extract the spatial distribution of surface potential across the ferroelectric domains. As shown in Fig. 1e, a contact potential difference ($V_{CPD}$) of approximately 70 mV is observed between the $P_{up}$ and $P_{down}$ regions (bottom curve). SKPM imaging of the ML $WS_2$ atop the poled AlScN (Fig. 1f) revealed ferroelectric-induced charge traps in ML $WS_2$. The inset zoom-in (on the white dotted region) shows a $V_{CPD}$ variation of ~60 mV across the three polarization regions (bottom curve). This result demonstrates that ferroelectric polarization can significantly alter the surface potential and work function of ML $WS_2$, thereby influencing its electrical and optical behavior efficiently. The ~10 meV discrepancy may result from imperfect interfacial contact between ML $WS_2$ and AlScN. Additionally, PFM and SKPM mappings for two additional patterned AlScN are conducted (Fig. S3), indicating reliable and stable polarization characteristics of AlScN.

To assess the modulation of optical dispersion, we first perform PL measurements on ML $WS_2$ across regions with different ferroelectric polarization states (Fig. 1h). A pronounced peak redshift and intensity reduction in PL are observed from $P_{up}$ to $P_{down}$ regions, indicating that the excitonic behavior of ML $WS_2$ can be effectively tuned by the ferroelectric polarization reversal. This observation aligns well with prior reports on ferroelectricity-tunable PL emission[18, 22, 28]. To spatially resolve this modulation, PL mapping (Fig. S4) is recorded on the same ML $WS_2$/AlScN device and the extracted PL peak positions from different polarization regions are shown in Fig. 1i, clearly illustrating a redshift in the exciton peak from up to down and a corresponding blueshift from down to up. We also show the Gaussian-fitted distribution of PL peak positions (Fig. S8d). These spatial variations correlate well with the underlying ferroelectric domain pattern, confirming the strong influence of ferroelectric polarization on excitonic properties in ML $WS_2$. A more detailed discussion of the underlying physical mechanism is provided in a later section. This significant variation in the excitonic peak demonstrates the potential for nonvolatile tunability of the optical dispersion in ML $WS_2$.

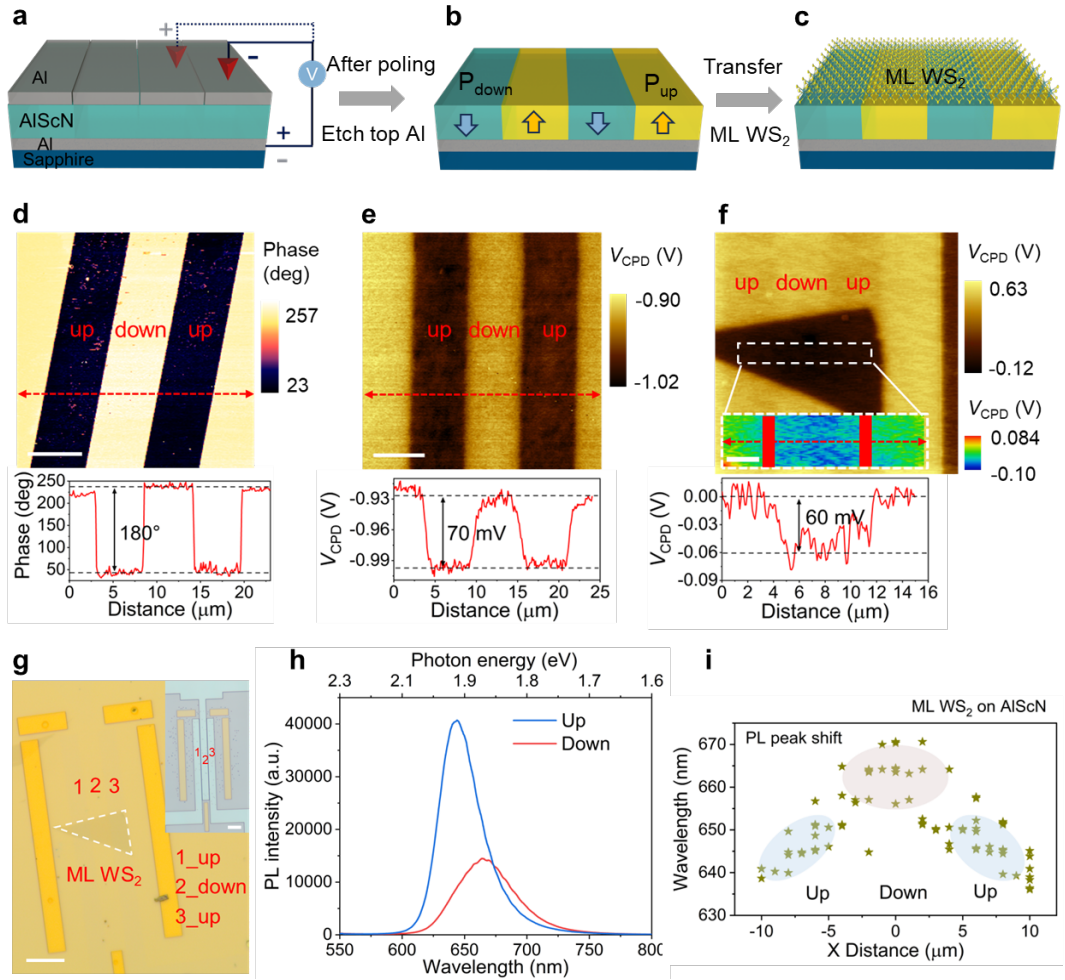

**Fig. 1 | Characterization of ML WS$_2$/AlScN device. a,** Schematic of AlScN ferroelectric domains locally polarized up and down by applying opposite voltages to the separated Al electrodes. **b,** Schematic of the AlScN pattern after poling and etching Al electrodes. Upward and downward arrows represent AlScN poling up (P$_{up}$) and down (P$_{down}$) states, respectively. **c,** Schematic of the ML WS$_2$/AlScN device after transferring ML WS$_2$ onto the patterned AlScN surface with defined polarization domains. **d,** PFM image of the poled AlScN surface. Regions labeled "up" and "down" correspond to P$_{up}$ and P$_{down}$ states, respectively. The bottom curve shows the phase profile along the red dashed line, indicating a 180° phase difference between domains. **e,** SKPM image of the poled AlScN surface. The bottom line profile shows a ~70 meV potential difference along the red dashed line on poled AlScN surface. **f,** SKPM image taken of ML WS$_2$ on the patterned AlScN. The inset shows a zoomed-in view of the white dashed box. The bottom line profile reveals a ~60 meV potential difference across the red dashed line on the WS$_2$ surface. Scale bars, 5 μm. **g,** Optical micrograph of the WS$_2$/AlScN device. The triangular transparent flake is the transferred ML WS$_2$ on the poled AlScN surface. Yellow rectangles are Au alignment markers. Scale bar, 10 μm. **h,** PL spectra acquired from Regions 2 (up) and 3 (down), showing a clear redshift and intensity variation. **i.** Extracted PL peak positions across Regions 1 (up), 2 (down), and 3 (up), illustrating the energy shift in response to ferroelectric polarization, as obtained from PL mapping.

**Imaging Spectroscopic Ellipsometry measurement of the ML WS$_2$/AlScN device**

The nonvolatile tunability of optical dispersion in ML WS$_2$, driven by ferroelectric polarization, can be effectively probed using ISE. Figure 2a schematically illustrates the SE measurement configuration, where linearly polarized light is incident on the sample and undergoes a polarization change to elliptical upon reflection. The changes in amplitude ($\psi$) and phase ($\Delta$) between parallel (p) and perpendicular (s) components of the reflected polarized light reveal the optical dispersion characteristics of sample, described by the complex reflection coefficient ratio: $\frac{r_p}{r_s} = \tan(\psi)\, e^{i\Delta}$.

Figures 2b and 2e show ISE mapping images of $\psi$ and $\Delta$ of ML WS$_2$ on the patterned AlScN acquired at an incidence angle of 65° (the mapping of $\psi$ with incidence angles of 70° and 75° are shown in Fig. S5) and a wavelength of 636 nm, near the A exciton resonance of ML WS$_2$. A distinct contrast is observed between regions with different polarization, indicating substantial tunability of the A exciton under the influence of the ferroelectric polarization. The full spectra of $\psi$ and $\Delta$, measured from 400 to 1000 nm at an incidence angle of 65°, are shown in Figs. 2c and 2f. A noticeable variation near the A exciton confirms that the excitonic response is modulated by the underlying ferroelectric domains. Magnified views of the 600–700 nm range are provided in the insets to highlight these differences more clearly. Notably, the contrast in $\psi$ is more pronounced than in $\Delta$ in the mapping images. This difference arises from $\psi$ being more sensitive to the absorption of the sample which is related the imaginary part ($k$) of the refractive index, and $k$ is modulated most strongly at the exciton resonance. Alternatively, $\Delta$ depends on the phase shift of reflected light which is primarily influenced by the real part ($n$), and $n$ is most strongly modulated at energies below the exciton resonance. To validate this, simulations of ($\psi$, $\Delta$) are conducted by the transfer matrix method (TMM), as presented in Figs. 2d and 2g. The simulation details are provided in the Supplementary. Similarly, both $\psi$ and $\Delta$ present distinct variations near the A exciton, with $\psi$ exhibiting a stronger response, indicating it is the dominant factor to exciton-related optical dispersion modulation. Additionally, it should be noted that the experiment results show some difference in $\psi$ and $\Delta$ near the B exciton (~530 nm). However, simulations indicate only a minor variation in $\psi$ and nearly no change in $\Delta$ in this region. This discrepancy is likely due to substrate effects, specifically, polarization-dependent reflectance changes in AlScN, which may enhance the differential reflectance around the B exciton. We also present both experimental and simulated ($\psi$, $\Delta$) spectra at incidence angles of 70° and 75° (Fig. S6), which show consistent trends with those observed at 65°.

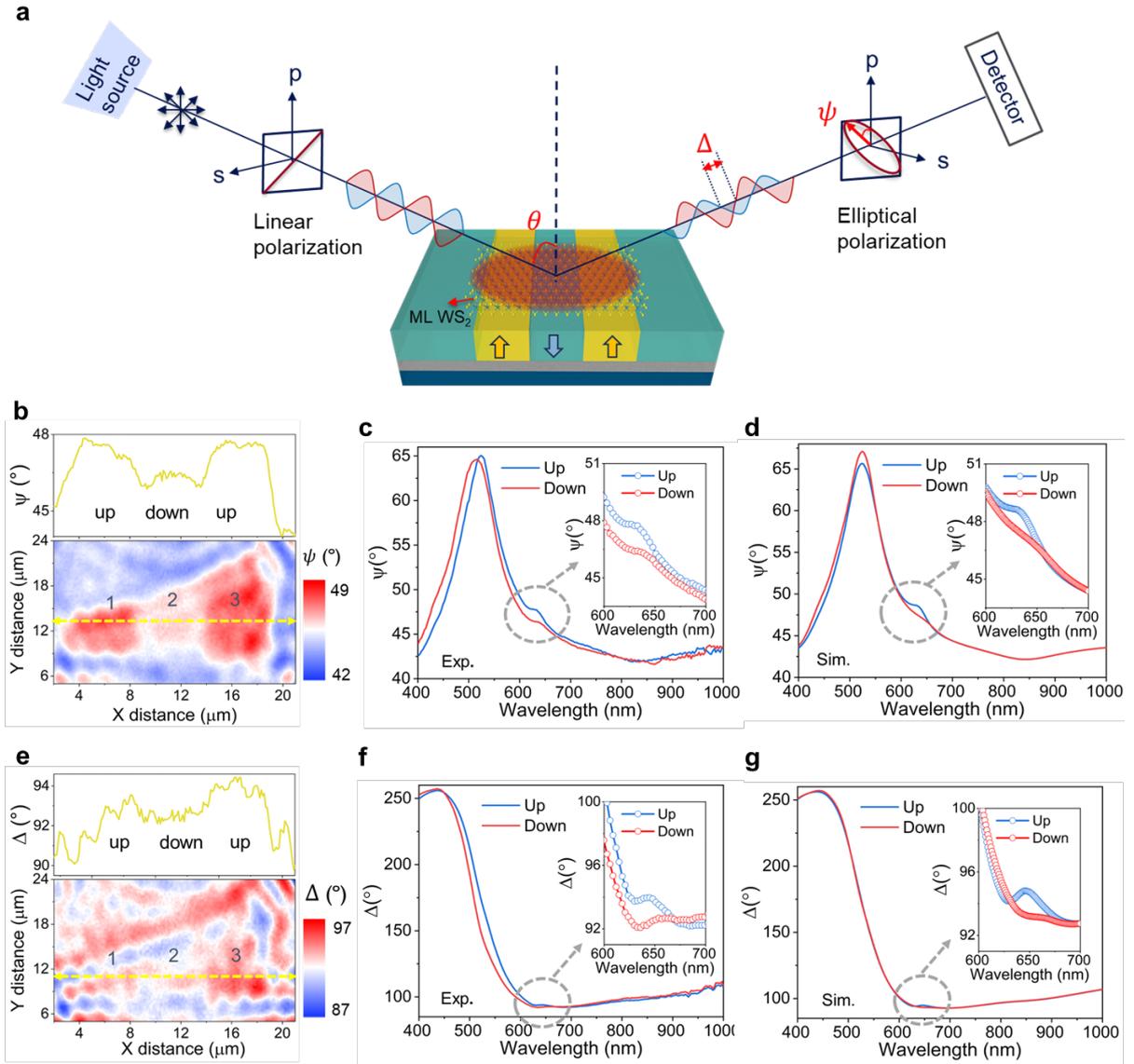

**Fig. 2 | Schematic illustration of the ISE measurement setup and SE measurement of ML WS$_2$/AlScN device. a.** Schematic of the ISE measurement configuration. Linearly polarized light is incident on the ML WS$_2$/AlScN surface at a determined angle of incidence (AOI: $\theta$). The detector measures the changes in amplitude ($\psi$) and phase ($\Delta$) between parallel (p) and perpendicular (s) components of the reflected elliptically polarized light. **b,e.** SE mapping of $\psi$ (**b**) and $\Delta$ (**e**) of ML WS$_2$ on patterned AlScN surface at the wavelength of 636 nm (around A exciton), AIO: 65°. The top profiles correspond to the yellow dashed lines in the respective mappings. Regions 1, 2, and 3 correspond to P$_{up}$, P$_{down}$, and P$_{up}$ states, respectively. The measured $\psi$ ($\Delta$) (**c,f**) and simulated $\psi$ ($\Delta$) (**d,g**) spectra from 400 to 1000 nm at an incidence angle of 65°. The insets show magnified views of the spectral range from 600 to 700 nm, highlighting features near the A exciton.

**Non-volatile optical dispersion tunability of ML WS$_2$/AlScN device**

The ISE data is acquired at multiple angles of incidence from 65° to 75° with 5° steps. A model-based fitting approach is employed to extract the optical properties of the multilayer stack (ML WS$_2$/AlScN/Al/Sapphire). To accurately model the optical dispersion of ML WS$_2$, the Tauc-Lorentz (TL) oscillator model[29] is adopted. This physical dispersion model effectively captures the key characteristics of excitonic transitions. Additional details of the fitting procedure are provided in the Methods and Fig. S7.

The extracted complex refractive indices ($n$, $k$) and dielectric permittivities ($\varepsilon_1$, $\varepsilon_2$) of ML WS$_2$ under P$_{up}$ and P$_{down}$ polarization states are presented in Figs. 3a and 3b. Three typical excitonic peaks are observed, corresponding to the A, B, and C excitons of ML WS$_2$[30]. Here, we mainly focus on the strongest excitonic modulation of the A exciton, characterized by a significant redshift and a decrease in amplitude. Whereas the B (~2.35 eV) and C (~2.70 eV) excitons display only minor variations in amplitude. At energies near the A exciton, the maximum of the real part of the refractive index shifts from 1.91 eV to 1.86 eV, while the imaginary part maximum shifts from 1.97 eV to 1.93 eV as the ferroelectric polarization is switched from P$_{up}$ to P$_{down}$ state. Consistent with our previous analysis, the maximum tunability of the real ($n$) and imaginary ($k$) components occurs at slightly different energies, reflecting the distinct physical origins of their contributions to the optical response, and this observation is consistent with the Tauc-Lorentz model. The energy shifts are comparable to previous reports of electrically gated refractive index tuning[10, 31, 32]. However, conventional gate-tuning methods typically require ultra-high voltages (up to 100 V), which are not only high energy consuming but also volatile. In contrast, the ferroelectric approach offers nonvolatile, reversible tuning without the sustained power consumption. To further illustrate this effect, we show the normalized $\psi$ spectra in Fig. 3c across a range of incidence angles (40° to 75° in 5° increments) for both P$_{up}$ state (solid line) and P$_{down}$ state (transparent line). Across nearly all incidence angles, the P$_{up}$ state consistently produces excitonic features with higher energy and amplitude than the P$_{down}$ state. For a more intuitive visualization of the modulation amplitude, we plot the differences $n_{up}$-$n_{down}$ ($\Delta n$) and $k_{up}$-$k_{down}$ ($\Delta k$) in Fig. 3d. The modulation around the A exciton resonance reaches up to 0.7 for $\Delta n$ and 0.4 for $\Delta k$—remarkably large compared to those achieved by traditional gate-induced tuning methods[10, 31] (Fig. S8). To further validate the tunability, we perform reflectance measurements along with simulations using the TMM, as shown in Figs. 3e and 3f. The simulated reflectance spectra closely match the experimental results, particularly in the region of the A exciton, reinforcing the validity of the observed trends. Notably, the ferroelectric polarization in this system is both nonvolatile and reversible, enabling programmable and energy-efficient modulation of optical dispersion without any physical constraint of the gate electrode or continuous power supply (Table S2).

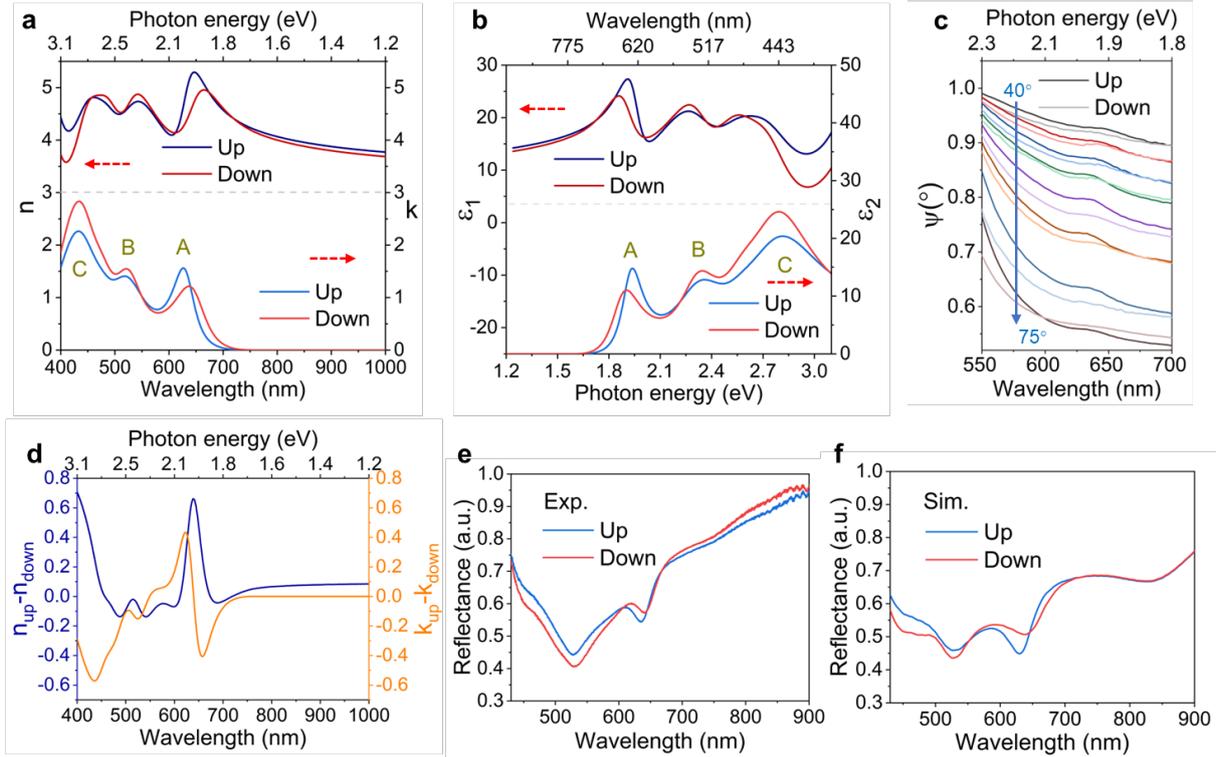

**Fig. 3 | Optical dispersion tunability of ML WS$_2$. a.** Fitted real part *n* (top) and imaginary part *k* (bottom) of the refractive index under different polarization states (P$_{up}$ and P$_{down}$). **b.** Fitted real part $\varepsilon_1$ (top) and imaginary part $\varepsilon_2$ (bottom) of the dielectric permittivity under P$_{up}$ and P$_{down}$ states. **c.** Normalized $\psi$ spectra measured at incidence angles from 40° to 75° in 5° steps for P$_{up}$ (solid lines) and P$_{down}$ (transparent lines), shown in the range of 550–700 nm (zoom-in of Fig. S6e), highlighting spectral differences near the A exciton resonance. **d.** Calculated tunable range of optical constants $n_{up}$-$n_{down}$ ($\Delta n$) and $k_{up}$-$k_{down}$ ($\Delta k$) between P$_{up}$ and P$_{down}$ states, demonstrating significant modulation near the A exciton. **e, f.** Experimental **(e)** and simulated **(f)** reflectance spectra of ML WS$_2$ under different polarization, showing consistent excitonic modulation behavior.

**Ferroelectric polarization induced asymmetric screening effect**

To gain insight into the physical process of the observed tunability in the complex refractive index, the excitonic effect in ML WS$_2$ on different substrates are analyzed in Supporting Information Section 1. We reasonably speculate that the bandgap of ML WS$_2$ undergoes renormalization caused by ferroelectric polarization, leading to a reduction of the quasiparticle bandgap in ML WS$_2$ on AlScN. Then additonal doping carriers in the AlScN system bind to neutral excitons to form trions, which further contribute to the redshift of the excitonic peak. Notably, the PL emission intensity of ML WS$_2$ on the AlScN P$_{up}$ state is the strongest (Fig. S9a), indicating enhanced populations of both neutral excitons and trions. However, under the P$_{down}$ state, a further redshift is observed along with a noticeable reduction in intensity compared to the P$_{up}$ state. To exclude strain effect as the cause of the redshift from P$_{up}$ to P$_{down}$, Raman spectra of ML WS$_2$ on different substrates are analyzed (Fig. S11). The E$_{2g}$ peak exhibits a redshift relative to SiO$_2$ and sapphire, suggesting strain plays a role in the overall excitonic redshift on AlScN[12]. However, the absence of a shift between the $P_{up}$ and $P_{down}$ states confirms that strain does not contribute to the polarization-dependent redshift. This raises an important question: what additional physical mechanism accounts for the more pronounced excitonic changes between the $P_{up}$ and $P_{down}$ states?

We ascribe this to asymmetric carrier screening in ML $WS_2$ under opposite ferroelectric polarizations. As-exfoliated ML $WS_2$ is typically n-type [Fig. 4b(i)] due to the chemical heterogeneity of S vacancies[33] that serve as donor states or shallow traps near the conduction band. The $P_{up}$ polarization of AlScN film induces electron doping shifting the Fermi level further into the conduction band [Fig. 4b(ii)], which substantially increases the local density of states (DOS) and thereby strengthens the screening effect ($E_{sc}$). Conversely, the $P_{down}$ polarization drives hole doping, but the DOS near the valence band is inherently lower [Fig. 4b(iii)] resulting in much weaker screening. Consequently, ML $WS_2$ on the $P_{down}$ region experiences stronger Coulomb interactions [Fig. 4a(ii) and (iii)] leading to an enhanced formation of positive trions and larger binding energies compared to the $P_{up}$ state, as schematically illustrated in Fig. 4c. Moreover, the shallow-traps become filled by the electron doping in $P_{up}$ state, which suppresses nonradiative recombination. This is comfirmed by the low-temperature (77 K) PL measurements, exhibiting negligible shift in $P_{up}$ state (Fig. S12b) whereas a clear blue-shift (Fig. S12c) in $P_{down}$ state compared to room temperature, and the decrease of intensity difference between the $P_{up}$ and $P_{down}$ states (Fig. S12a). The relative densities of neutral excitons and trions depend on the density of unbound carriers at the Fermi level, which can be estimated using the mass-action law[5, 8, 11], $\frac{n_{X^0} n_e}{n_{X^-}} = A k_B T exp(-\frac{E_{X^-}^b}{k_B T})$, where $n_{X^0}$, $n_{X^-}$ and $n_e$ are the densities of neutral excitons, trions, and unbound electrons, respectively. The derived unbound carrier densities are approximately $1.313 \times 10^{12}\,cm^{-2}$ for the $P_{up}$ state and $5.231 \times 10^{12}\,cm^{-2}$ for the $P_{down}$ state (see Supporting Information section 3 for details). That means the hole doping is more effective under the $P_{down}$ state because intrinsically n-type $WS_2$ experiences a stronger screening effect in the $P_{up}$ state, making electron injection more difficult[19]. This is further supported by the PL mapping of wet-transferred MOCVD-grown ML $WS_2$, which clearly shows the enhancement in the $P_{up}$ state (Fig. 4d,e and Fig. S14). Usually more unbound carriers along with higher probabilities for the formation of trions[5, 8] that supports the proposed asymmetric screening model. Also, this indicates that doping achievable in the AlScN ferroelectric system outperforms that of the majority of reported chemical and alternative ferroelectric doping junctions[3, 18, 23, 34-36].

To access the intuitive physical process, we show in Fig. S15 a schematic evolution of the electronic and excitonic properties of ML $WS_2$ under different situations. The transition from $SiO_2$ to AlScN $P_{up}$ is expected to decrease both the quasiparticle bandgap ($E_{qg}$) and the exciton binding energies ($E_{X^0}^b$, $E_{X^-}^b$) [Fig. S15b (i), (ii)], accompanied by increased densities of neutral excitons ($X^0$) and negative trions ($X^-$) [Fig. S15a (i), (ii)]. The transition from the AlScN $P_{up}$ to $P_{down}$ is expected to substantially increase the positive trion ($X^+$) density [Fig. S15a (iii)], as well as increase the binding energies of both neutral excitons and trions ($E_{X^+}^b$) [Fig. S15b (iii)], arising from asymmetric carrier screening.

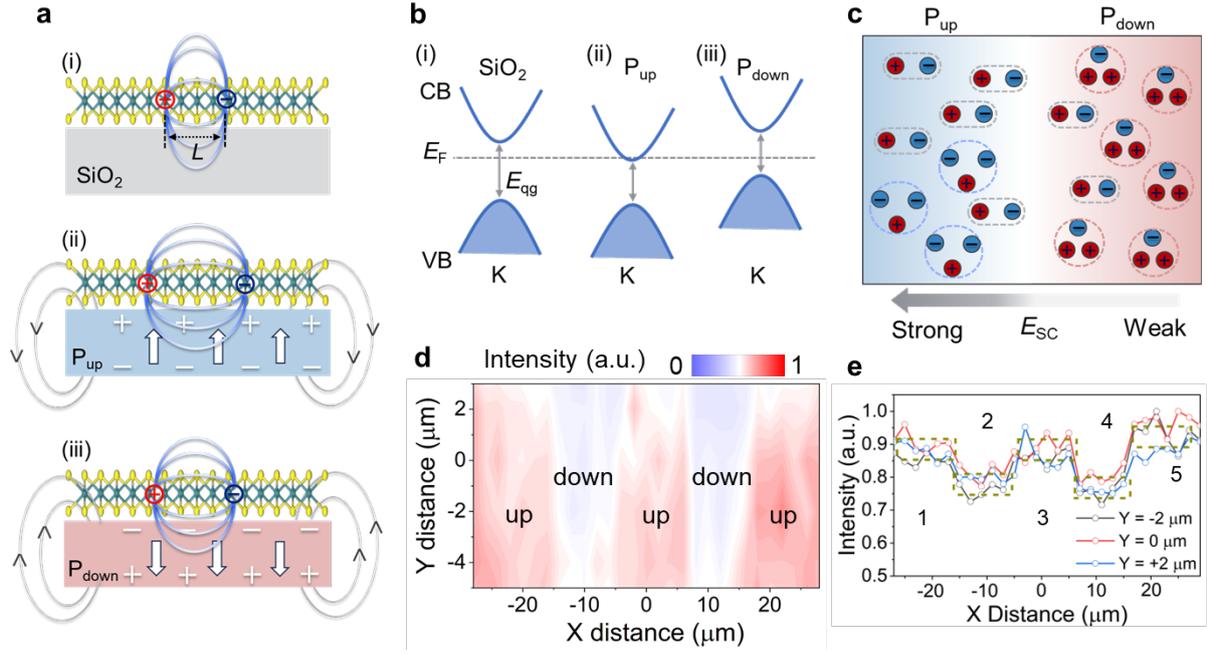

**Fig. 4 | Asymmetric screening effect in ML WS$_2$. a.** Real-space schematic of electron and hole bound into exciton in ML WS$_2$ supported on different substrates: (i) SiO$_2$, (ii) AlScN P$_{up}$ state, and (iii) AlScN P$_{down}$ state. **b.** Band diagrams of ML WS$_2$ on SiO$_2$, AlScN P$_{up}$, and AlScN P$_{down}$ substrates, respectively. The conduction band (CB) and valence band (VB) of ML WS$_2$ are shown relative to the Fermi level $E_F$ under different doping configurations. The gray dashed line denotes the Fermi level position, and the gray arrows denote the quasiparticle bandgap $E_{qg}$. **c.** Asymmetric carrier screening in ML WS$_2$ induced by opposite ferroelectric polarization. The P$_{down}$ region exhibits a stronger binding between charges than the P$_{up}$ region due to weaker screening, resulting in higher trion density. $E_{sc}$ represents the screening field arising from polarization-induced carrier doping. **d.** Normalized spatial PL mapping of wet-transferred MOCVD-grown ML WS$_2$ on patterned AlScN surface. **e.** The profiles of Y = -2, 0, and +2 µm extracted from **d**. Labels 1, 3, and 5 correspond to P$_{up}$ states, and Labels 2 and 4 correspond to P$_{down}$ states.

## Electrical observation of ML WS$_2$ n-p homojunction

To further explore the potential n-p homojunction formation induced by the oppositely polarized AlScN domains, we investigate the electrical properties of the ML WS$_2$/AlScN device. Building on the ML WS$_2$/AlScN/Al/Sapphire structure used in optical measurements, two kinds of contact metals, In/Au and Pd, are deposited onto the electron-doped (regarded as n-type) and hole-doped (regarded as p-type) regions of ML WS$_2$, respectively (see Methods for details). Figure 5a illustrates the schematic of the device fabrication process. Figure 5b presents the PFM phase contrast of AlScN under two oppositely poled domains (phase profile shown in Fig. S16a). An optical micrograph of the ML WS$_2$ atop the patterned AlScN domains is also shown, with the transferred ML WS$_2$ flake outlined by the white dashed circle. As discussed earlier, the P$_{up}$ domain induces electron doping in ML WS$_2$, while the P$_{down}$ domain induces hole doping, enabling band structure modulation and forming a lateral n-p homojunction across the ferroelectric domain boundary. The current-voltage (I-V) characteristics of the ML WS$_2$ n-p homojunction, modulated by the underlying ferroelectric polarization, are shown in Fig. 5c. The device exhibits strong rectifying behavior, with a forward threshold voltage around 0.8 V. The insets in Fig. 5c show schematic band diagrams of the ML $WS_2$ n-p homojunction under forward and reverse bias. To further analyze the rectification, Figure 5d plots the I-V curve of the n-p configuration on a semi-logarithmic scale. The rectification ratio at ±3 V is calculated to be $R_{np} = 6×10^3$. The I-V curves of the n-p configuration with different scanned voltages is shown in Fig. S16b. For comparison, the rectification ratios of the n-n and p-p configurations are $R_{nn} = 20$ and $R_{pp} = 10$, respectively (Fig. S16c). The nearly symmetric behavior of the n-n and p-p curves confirm that the observed rectification arises from the n-p homojunction, rather than from small Schottky barriers at the metal interfaces. In the forward bias regime, the I–V curve of the n-p junction can be well-fitted with the Shockley diode equation[37], yielding an ideality factor of ~2.9. This value suggests that carrier recombination dominates at the junction, and a relatively good-quality n-p junction interface. The rectification ratio and ideality factor of our device are comparable to those of volatile homojunctions formed via split-gate biasing[38, 39]. These electrical measurements offer clear evidence of ferroelectric-induced modulation of carrier type in ML $WS_2$.

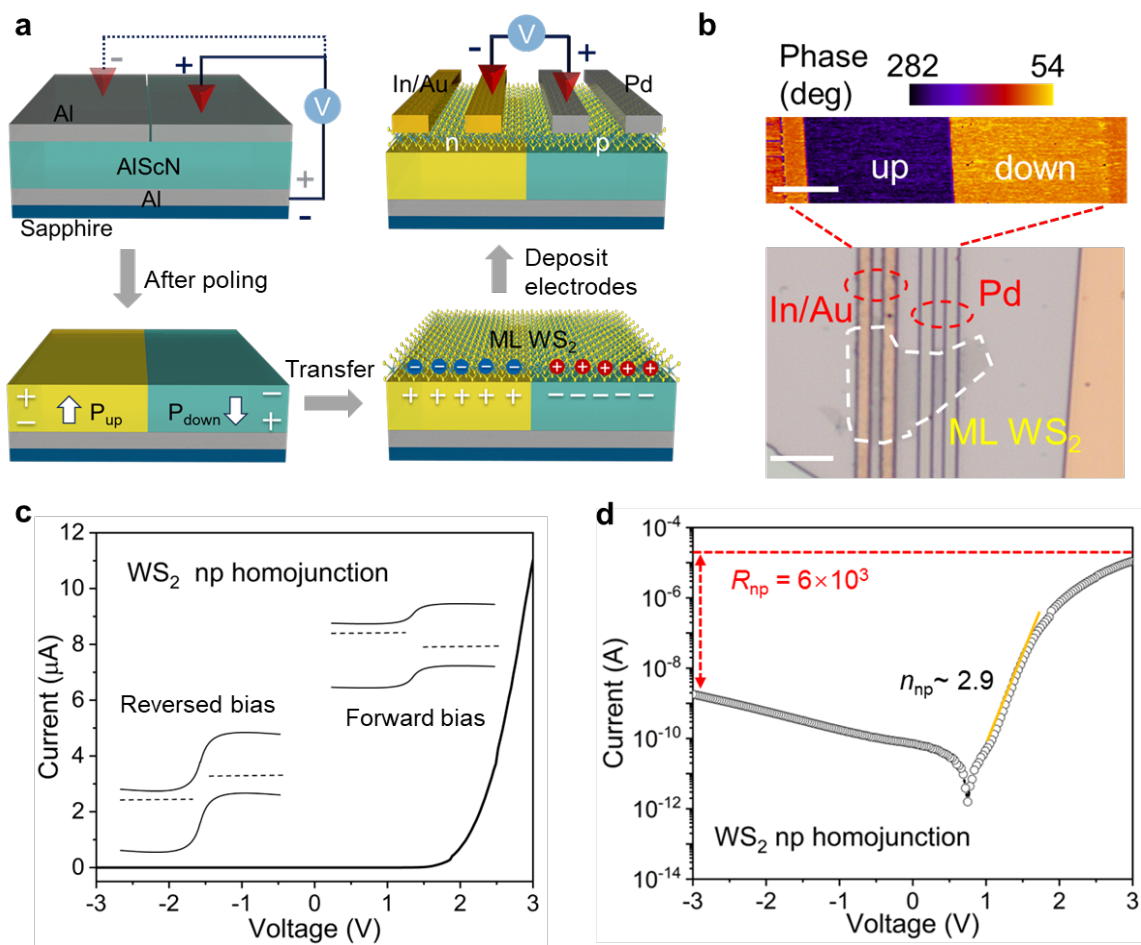

**Fig. 5 | Ferroelectric induced the ML $WS_2$ n-p homojunction. a.** Schematic illustration of the device fabrication process: AlScN electrical poling, patterned AlScN after poling, ML $WS_2$/AlScN after transferring, and final device with In/Au contacts on n-doped ML $WS_2$ and Pd contacts on p-doped ML $WS_2$. Up and down arrows denote AlScN $P_{up}$ and $P_{down}$ states, respectively. **b.** Top: PFM image showing

a 180° phase difference between AlScN $P_{up}$ and $P_{down}$ domains. Scale bar: 5 μm. Bottom: Optical micrograph for the ML WS$_2$/AlScN device. The white dashed circle indicates the transferred ML WS$_2$ flake on $P_{up}$ and $P_{down}$ regions. Scale bar, 10 μm. **c.** I–V characteristic of the ML WS$_2$/AlScN device, showing rectifying behavior characteristic of a homojunction. Insets: schematic band diagrams of the ML WS$_2$ n-p homojunction under forward and reverse bias. **d.** Semi-log plot of the I–V curve for the n-p homojunction, with a Shockley diode model fit shown in yellow. The extracted ideality factor is $n_{np}$ ~ 2.9, and the rectification ratio at $V = \pm 3$ V is $R_{np} = 6 \times 10^3$.

## Discussion:

We have demonstrated robust nonvolatile control over optical dispersion in monolayer WS$_2$ through engineered ferroelectric domains in AlScN. Our approach delivers record-high modulation amplitudes ($\Delta n > 0.7$, $\Delta k > 0.4$) without continuous power consumption, a performance previously only attained through electrostatic gate-tuning with voltages approaching 100 V. The discovered asymmetric screening mechanism provides fundamental insights into how ferroelectric polarization governs excitonic behavior through carrier-density-dependent Coulomb interactions.

The demonstrated gate-free p-n homojunction, with a rectification ratio of $6 \times 10^3$, validates the effectiveness of ferroelectric domain engineering for creating functional electronic devices. This achievement, combined with the nonvolatile optical modulation, establishes a new paradigm for integrated photonic and optoelectronic systems.

Looking forward, this work opens several compelling directions. The spatial resolution of ferroelectric domains could be pushed to nanoscale dimensions using advanced poling techniques, enabling metasurface applications with programmable optical responses. Integration with photonic waveguides could yield nonvolatile optical switches and modulators with ultralow power consumption. Furthermore, the asymmetric screening model suggests opportunities for engineering novel excitonic states and many-body phenomena through precise polarization control.

## Methods

### AlScN deposition

The fabrication process commenced with the in situ deposition of a 50 nm Al seed layer at 150 °C under a 20 sccm argon (Ar) atmosphere, using 900 W RF power, on a 6-inch c-plane sapphire wafer (0001) substrate. The 45 nm Al$_{0.64}$Sc$_{0.36}$N layer was then co-sputtered at 350 °C using 900 W and 700 W RF power applied to 100 mm Al and Sc targets, respectively, under a 30 sccm nitrogen (N$_2$) atmosphere. A 50 nm Al capping layer was subsequently deposited under the same conditions as the Al seed layer. All layers were deposited sequentially in an Evatec Clusterline 200 II system without vacuum break to suppress oxidation and preserve film quality.

### Device fabrication

The fabrication process began with a 1% hydrofluoric acid (HF) wet etch to remove the top aluminum capping layer. The first electron beam lithography (EBL) step was then performed using an EBPG5200+ (Raith) to define alignment markers, followed by lift-off of evaporated 5 nm Ti/ 60 nm Au. A second EBL step was used to pattern the tuning electrodes for ferroelectric modulation. Subsequently, 60 nm of high-purity Al was deposited using an Intlvac electron-beam evaporator, followed by lift-off. After electrical poling, the Al electrodes were selectively removed by wet etching (1% HF), while preserving the Au alignment markers. The ML $WS_2$ flakes were mechanically exfoliated from a bulk crystal onto a $SiO_2$/Si substrate with a 285 nm thick $SiO_2$. The exfoliated flakes were then transferred onto the electrically poled AlScN substrate using the pick-up transfer technique with a polydimethylsiloxane (PDMS) stamp coated with a polycarbonate (PC) film[40]. The growth details of MOCVD-grown ML $WS_2$ and the wet-transfer processes follow standard procedures described in previous report[31]. For diode fabrication, two additional EBL steps were employed to define the contacts. 10 nm Indium/ 40 nm Au was deposited via thermal evaporation using a Lesker Nano-36 system, and 60 nm of Pd was deposited on the opposite electrode using a Lesker PVD75 electron-beam evaporator, both followed by standard lift-off processes.

**PFM and SKPM characterization**

PFM and SKPM measurements were performed on a commercial Asylum Research system using Pt/Ir-coated silicon probes operating at a contact-resonance frequency of 75 kHz.

**Optical measurements**

Reflectance, PL, and Raman spectra were conducted using a Horiba Scientific confocal microscope (LabRAM HR Evolution) equipped with an Olympus objective lens (up to ×100) and three spectrometers (100, 600, and 1800 gr/mm gratings) coupled to a Si focal plane array detector. For reflectance measurements, the reflected intensity from a polished Ag mirror was used as a reference to normalize the sample signals. PL spectra were obtained using a 532 nm continuous-wave laser with a ×100 objective at 0.1% laser power (240 μW). Raman spectra were recorded using the same 532 nm excitation at 1% laser power (1.57 mW). A linear polarizer was placed in the excitation path prior to the sample. For low-temperature PL measurements at 77 K, the sample was mounted in a Linkam cryostat and cooled by controlled liquid nitrogen flow.

**Imaging spectroscopic ellipsometry**

Spectroscopic ellipsometry measurements were carried out using a Parks System Ellipsometer (Accurion EP4) in Delta & Psi (standard RCE) mode, covering the ultraviolet to near-infrared range (190-1000 nm) and variable incidence angles (40°-80°). Data were analyzed using the EP4Model software, employing a multi-Tauc–Lorentz oscillator model to extract the complex optical constants. Mapping images were extracted using DataStudio software. A ×50 objective lens provided a lateral resolution of ~1 μm. The measured sample consisted of a multilayer stack: ML $WS_2$ (0.7 nm)/AlScN (45 nm)/Al (50 nm)/Sapphire.

**Simulations**

The transfer matrix method (TMM) was employed using open-source code to simulate the reflectance and ellipsometric parameters ($\psi$, $\Delta$) spectra of ML WS$_2$ under different AlScN polarization states[41-43] The multilayer model consisted of ML WS$_2$ (0.7 nm)/AlScN (45 nm)/Al (50 nm)/Sapphire. The complex refractive indices used in the simulations were experimentally obtained via spectroscopic ellipsometry. To extract ($\psi$, $\Delta$), Fresnel reflection coefficients ($r_s$ and $r_p$) for s- and p-polarized light were calculated, from which the ellipsometric parameters and corresponding reflectance were derived. The refractive indices of Al[44] and sapphire[45] were adopted from literature values.

**Electrical measurements**

AlScN electrical poling and ML WS$_2$ homojunction measurements were performed at room temperature using a probe station integrated with a Keithley 4200A-SCS semiconductor parameter analyzer (Tektronix Inc.), equipped with the non-volatile memory (NVM) test library.

**References**

1. Lopez-Sanchez, O., Lembke, D., Kayci, M., Radenovic, A. & Kis, A. Ultrasensitive photodetectors based on monolayer MoS$_2$. *Nat. Nanotechnol.* **8**, 497–501 (2013).
2. Thureja, D. et al. Electrically tunable quantum confinement of neutral excitons. *Nature* **606**, 298–304 (2022).
3. Ming-Yang Li, Y.S., Chia-Chin Cheng, Li-Syuan Lu, Yung-Chang Lin, Hao-Lin Tang, Meng-Lin Tsai, Chih-Wei Chu, Kung-Hwa Wei Epitaxial growth of a monolayer WSe$_2$-MoS$_2$ lateral p-n junction with an atomically sharp interface. *Science* **349**, 524–528 (2015).
4. Chernikov, A. et al. Exciton binding energy and nonhydrogenic Rydberg series in monolayer WS$_2$. *Phys. Rev. Lett.* **113**, 076802 (2014).
5. Rahul Kesarwani, K.B.S., Teng-De Huang, Yu-Fan Chiang, Nai-Chang Yeh, Yann-Wen Lan, Ting-Hua Lu Control of trion-to-exciton conversion in monolayer WS$_2$ by orbital angular momentum of light. *Sci. Adv.* **8**, eabm0100 (2022).
6. Torche, A. & Bester, G. First-principles many-body theory for charged and neutral excitations: Trion fine structure splitting in transition metal dichalcogenides. *Phys. Rev. B* **100**, 201403(R) (2019).
7. Mitioglu, A.A. et al. Optical manipulation of the exciton charge state in single-layer tungsten disulfide. *Phys. Rev. B* **88**, 245403 (2013).
8. Mak, K.F. et al. Tightly bound trions in monolayer MoS$_2$. *Nat. Mater.* **12**, 207–211 (2013).
9. Ross, J.S. et al. Electrical control of neutral and charged excitons in a monolayer semiconductor. *Nat. Commun.* **4**, 1474 (2013).
10. Yu, Y. et al. Giant Gating Tunability of Optical Refractive Index in Transition Metal Dichalcogenide Monolayers. *Nano Lett.* **17**, 3613–3618 (2017).
11. Harats, M.G., Kirchhof, J.N., Qiao, M., Greben, K. & Bolotin, K.I. Dynamics and efficient conversion of excitons to trions in non-uniformly strained monolayer WS$_2$. *Nat. Photonics* **14**, 324–329 (2020).
12. Kim, G. et al. High-Density, Localized Quantum Emitters in Strained 2D Semiconductors.

ACS Nano **16**, 9651–9659 (2022).

13. Absor, M.A.U., Kotaka, H., Ishii, F. & Saito, M. Strain-controlled spin splitting in the conduction band of monolayer $WS_2$. *Phys. Rev. B* **94**, 115131 (2016).
14. Stier, A.V., McCreary, K.M., Jonker, B.T., Kono, J. & Crooker, S.A. Exciton diamagnetic shifts and valley Zeeman effects in monolayer $WS_2$ and $MoS_2$ to 65 Tesla. *Nat. Commun.* **7**, 10643 (2016).
15. Steinhoff, A. et al. Biexciton fine structure in monolayer transition metal dichalcogenides. *Nat. Phys.* **14**, 1199–1204 (2018).
16. Kang, Y. et al. Spatially selective p-type doping for constructing lateral $WS_2$ p-n homojunction via low-energy nitrogen ion implantation. *Light Sci. Appl.* **13**, 127 (2024).
17. Wu, G. et al. Ferroelectric-defined reconfigurable homojunctions for in-memory sensing and computing. *Nat. Mater.* **22**, 1499–1506 (2023).
18. Wu, G. et al. Programmable transition metal dichalcogenide homojunctions controlled by nonvolatile ferroelectric domains. *Nat. Electron.* **3**, 43–50 (2020).
19. Chen, J.W. et al. A gate-free monolayer $WSe_2$ pn diode. *Nat. Commun.* **9**, 3143 (2018).
20. Li, D. et al. Giant Transport Anisotropy in $ReS_2$ Revealed via Nanoscale Conducting-Path Control. *Phys. Rev. Lett.* **127**, 136803 (2021).
21. Singh, S. et al. Nonvolatile Control of Valley Polarized Emission in 2D $WSe_2$-AlScN Heterostructures. *ACS Nano* **18**, 17958–17968 (2024).
22. Li, C.H., McCreary, K.M. & Jonker, B.T. Spatial Control of Photoluminescence at Room Temperature by Ferroelectric Domains in Monolayer $WS_2$/PZT Hybrid Structures. *ACS Omega* **1**, 1075–1080 (2016).
23. Wen, B. et al. Ferroelectric-Driven Exciton and Trion Modulation in Monolayer Molybdenum and Tungsten Diselenides. *ACS Nano* **13**, 5335–5343 (2019).
24. Kim, K.H., Karpov, I., Olsson, R.H., 3rd & Jariwala, D. Wurtzite and fluorite ferroelectric materials for electronic memory. *Nat. Nanotechnol.* **18**, 422–441 (2023).
25. Kim, K.H. et al. Scalable CMOS back-end-of-line-compatible AlScN/two-dimensional channel ferroelectric field-effect transistors. *Nat. Nanotechnol.* **18**, 1044–1050 (2023).
26. Dhiren K. Pradhan, D.C.M., Gwangwoo Kim, Yunfei He, Pariasadat Musavigharavi, Kwan-Ho Kim, Nishant Sharma, Zirun Han, Xingyu Du, Venkata S. Puli, Eric A. Stach, W. Joshua Kennedy, Nicholas R. Glavin, Roy H. Olsson III & Deep Jariwala A scalable ferroelectric non-volatile memory operating at 600 °C. *Nat. Electron.* **7**, 348–355 (2024).
27. Tang, Z., Esteves, G. & Olsson, R.H. Sub-quarter micrometer periodically poled $Al_{0.68}Sc_{0.32}N$ for ultra-wideband photonics and acoustic devices. *J. Appl. Phys.* **134**, 114101 (2023).
28. Li, X. et al. Nonvolatile Electrical Valley Manipulation in $WS_2$ by Ferroelectric Gating. *ACS Nano* **16**, 20598–20606 (2022).
29. Hilfiker, J.N. & Tiwald, T. in Spectroscopic Ellipsometry for Photovoltaics: Fundamental Principles and Solar Cell Characterization, Vol. 1. (eds. H. Fujiwara & R.W. Collins) 115–153 (Springer International Publishing, Cham; 2018).
30. Liu, H.-L. et al. Optical properties of monolayer transition metal dichalcogenides probed by spectroscopic ellipsometry. *Appl. Phys. Lett.* **105**, 201905 (2014).
31. Lynch, J. et al. Full $2\pi$ phase modulation using exciton-polaritons in a two-dimensional superlattice. *Device* **3**, 100639 (2025).
32. Wang, Z. et al. Greatly Enhanced Resonant Exciton-Trion Conversion in Electrically

Modulated Atomically Thin WS$_2$ at Room Temperature. *Adv. Mater.* **35**, e2302248 (2023).
33. Lien, D.-H. et al. Electrical suppression of all nonradiative recombination pathways in monolayer semiconductors *Science* **364**, 468–471 (2019).
34. Lee, C.H. et al. Atomically thin p-n junctions with van der Waals heterointerfaces. *Nat. Nanotechnol.* **9**, 676–681 (2014).
35. Mao, X. et al. Nonvolatile Electric Control of Exciton Complexes in Monolayer MoSe$_2$ with Two-Dimensional Ferroelectric CuInP$_2$S$_6$. *ACS Appl. Mater. Interfaces* **13**, 24250–24257 (2021).
36. Xiao, Z., Song, J., Ferry, D.K., Ducharme, S. & Hong, X. Ferroelectric-Domain-Patterning-Controlled Schottky Junction State in Monolayer MoS$_2$. *Phys. Rev. Lett.* **118**, 236801 (2017).
37. Banwell, T.C. & Jayakumar, A. Exact analytical solution for current flow throughdiode with series resistance. *Electron. Lett.* **36**, 291–292 (2000).
38. Bie, Y.Q. et al. A MoTe$_2$-based light-emitting diode and photodetector for silicon photonic integrated circuits. *Nat. Nanotechnol.* **12**, 1124–1129 (2017).
39. Baugher, B.W., Churchill, H.O., Yang, Y. & Jarillo-Herrero, P. Optoelectronic devices based on electrically tunable p-n diodes in a monolayer dichalcogenide. *Nat. Nanotechnol.* **9**, 262–267 (2014).
40. Zomer, P.J., Guimarães, M.H.D., Brant, J.C., Tombros, N. & van Wees, B.J. Fast pick up technique for high quality heterostructures of bilayer graphene and hexagonal boron nitride. *Appl. Phys. Lett.* **105**, 013101 (2014).
41. Pettersson, L.A.A., Roman, L.S. & Inganäs, O. Modeling photocurrent action spectra of photovoltaic devices based on organic thin films. *J. Appl. Phys.* **86**, 487–496 (1999).
42. Peter Peumans, A.Y., Stephen R. Forrest Small molecular weight organic thin-film photodetectors and solar cells. *J. Appl. Phys.* **93**, 3693–3723 (2003).
43. Passler, N.C. & Paarmann, A. Generalized 4 × 4 matrix formalism for light propagation in anisotropic stratified media: study of surface phonon polaritons in polar dielectric heterostructures: erratum. *J. Opt. Soc. Am. B* **36**, 3246 (2019).
44. McPeak, K.M. et al. Plasmonic Films Can Easily Be Better: Rules and Recipes. *ACS Photonics* **2**, 326–333 (2015).
45. Malitson, I.H. Refraction and Dispersion of Synthetic Sapphire. *J. Opt. Soc. Am.* **52**, 1377–1379 (1962).

**Data availability**

The refractive index values are available in Supplementary Table S3. All other data are available upon request of the corresponding author.

**Acknowledgements**

D.J. and Y.C. acknowledge primary support from Office of Naval Research (ONR) Nanoscale Computing and Devices program (No. N00014−24−1−2131). Y.C. acknowledges the support of postdoctoral fellowship grant from Shanghai Jiao Tong University (SJTU grant). A portion of the sample fabrication, assembly, and characterization were carried out at the Singh Center for Nanotechnology at the University of Pennsylvania, which is supported by the National Science Foundation (NSF) National Nanotechnology Coordinated Infrastructure Program (Grant No. NNCI-1542153). The MOCVD $WS_2$ films were grown in the 2D Crystal Consortium–Materials Innovation Platform (2DCC-MIP) facility which is supported by the National Science Foundation under cooperative agreement DMR-2039351.

**Author contributions**

D.J. conceived the idea and supervised the project. D.J. and Y.C. designed the experiment. Y.C. performed the measurements and characterizations, and conducted the data fitting, with the help of J.L., B.C., Z.H. and H.C. Z.H. carried out the device fabrication processes with the help of H.C. and C.L. B.C. performed the TMM simulations. Y.C. exfoliated the ML $WS_2$ and K.M.Y carried out the $WS_2$ flakes transfer. C.C. and J.M.R. provided the MOCVD-grown $WS_2$ monolayers. Y.C. carried out the wet-transferred ML $WS_2$ with the help of J.L. D.J., Y.C. and Z.H. led the writing of the paper. Y.C. analyzed the data, built up the theoretical model, prepared the figures, and wrote the paper with the helpful discussion of the other authors.

**Competing interests**

The authors declare no competing interests.

Supplementary Information for

# Controlled non-volatile modulation of optical dispersion in monolayer tungsten disulfide via ferroelectric polarization patterning

Yuhong Cao[1], Zekun Hu[1], Jason Lynch[1], Bongjun Choi[1], Kyung Min Yang[1], Hyunmin Cho[1], Chloe Leblanc[1], Chen Chen[2], Joan M. Redwing[2], Deep Jariwala[1]*

[1]Department of Electrical and Systems Engineering, University of Pennsylvania, Philadelphia, Pennsylvania 19104, USA

[2]2D Crystal Consortium Materials Innovation Platform, Materials Research Institute, Penn State University, University Park, PA, USA

*Corresponding author: Deep Jariwala (email: dmj@seas.upenn.edu)

**Table of Contents**

**Section 1: Dielectric-screened Coulomb interaction and bandgap renormalization**

Previous works[1-3] have suggested that the dielectric environment strongly influences Coulomb screening interactions within the sheet, which in turn affects the exciton binding energies, that is dielectric-screened Coulomb interaction. In a 2D many-body system, the interaction potential between charged particles is given by[1,3], $V_{2D}(L) = \frac{e^2}{\varepsilon_0 \kappa L_0} f(\frac{L}{L_0})$, where $\kappa = (1 + \kappa_{\text{sub}})/2$ is the average dielectric constant of the surrounding media (air or vacuum, $\kappa = 1$, and substrate $\kappa_{\text{sub}}$), $f(\frac{L}{L_0})$ is a dimensionless screening function, $\varepsilon_0$ is the vacunm permittivity, $e$ is the electron charge, $L_0$ is the screening length, and $L$ is the electron–hole separation distance [smaller $L$ implies stronger binding, as shown in Fig. 4a(i)]. The exciton binding energies thus scales nonlinearly with $\kappa$ as, $E_X^b = E_b(\kappa = 1)/\kappa^{\alpha_X}$, where $E_b(\kappa = 1)$ is the exciton binding energy in vacuum, $\alpha_X$ is an empirical factor, and the subscript X refers to neutral exciton (X⁰), negative (X⁻), or positive (X⁺) trions. This inverse asymptotic relationship between $E_X^b$ and $\kappa$ is confirmed by both DFT simulations[1] and experiments[2,3]. Thus, we conduct the PL and reflectance measurements of ML WS$_2$ on SiO$_2$ ($\kappa$~3.9)[2] and sapphire ($\kappa$~10)[1], as shown in Fig. S9a-c. As expected, a blue-shift is observed from SiO$_2$ to sapphire, consistent with reduced binding energies due to stronger dielectric screening (Fig. S10a). However, ML WS$_2$ on AlScN ($\kappa$~16)[4] exhibits a red-shift, contrary to expectations based solely on binding energies. We also present the Gaussian-fitted PL peak distributions of ML WS$_2$ on SiO$_2$, AlScN P$_{\text{up}}$ and P$_{\text{down}}$ states (Fig. S9d), which further validate the observed trend from SiO$_2$ to P$_{\text{up}}$ to P$_{\text{down}}$.

Based on these results, we reasonablely speculate that the bandgap of ML WS$_2$ has undergone renormalization caused by ferroelectric polarization, resulting in the significant redshift. There are two pathways to achieve bandgap renormalization (BGR). One is the electron self-energy corrections, which are influenced by dielectric screening—an effect particularly pronounced in ML TMDs[3]. The renormalized quasiparticle bandgap can be expressed as, $E_{\text{qg}} = E_{\text{sp}} + \frac{E_{BGR}(\kappa=1)}{\kappa^\beta}$, where $E_{\text{sp}}$ is the single-particle bandgap, $E_{BGR}(\kappa = 1)$ is the BGR self-energy term in vacuum, and $\beta$ is a scaling factor. Similar to binding energies, $E_{\text{qg}}$ decreases with increasing $\kappa$. The second is carrier-induced BGR[5]. On the AlScN substrate, opposite polarizations introduce electron (P$_{\text{up}}$) and hole (P$_{\text{down}}$) doping, leading to a variation in background carrier density by nearly an order of magnitude (see Supporting Information Section 2 for details). This modulation level is comparable to that achieved by conventional chemical or elemental doping methods[6]. The combined effects of enhanced dielectric screening and ferroelectric-induced doping result in a substantial reduction of the quasiparticle bandgap in ML WS$_2$ on AlScN, more pronounced than that observed on sapphire, as shown in Fig. S10b and Fig. 4b.

## Section 2: Calculation of background carrier density variation induced by different AlScN polarization

Scanning Kelvin Probe Microscopy (SKPM) allows the estimation of the carrier density[7]. However, the actual Fermi-level is often inaccurate due to the influence of surface adsorbates. Hence, we focus on extracting the difference in carrier density between ML WS$_2$ regions on oppositely polarized AlScN domains. As shown by the line profile in Fig. 1f, a surface potential difference of ~60 mV is observed, which is comparable to values reported for chemically or elementally doped junctions[6]. Based on the carrier density equation,

$$n = N_C exp(-\frac{E_C - E_F}{K_B T})$$

Where, $N_C$ denotes the effective density of states in the conduction band, $T$ is the room temperature (300 K), and $k_B$ is the Boltzmann constant, with $K_B T \approx 25.9 \, meV$. According to the relationship between the work function ($\varphi$) and the contact potential difference ($V_{CPD}$) obtained from SKPM,

$$\varphi_{WS_2} = \varphi_{tip} - V_{CPD}$$

Where, $\varphi_{WS_2}$ and $\varphi_{tip}$ are the work function of ML WS$_2$ and tip, respectively. Then we can get the Fermi-level,

$$E_F = E_C - (\varphi_{WS_2} - \chi) = E_C - (\varphi_{tip} - V_{CPD} - \chi)$$

Where, $E_F$ and $E_C$ represent the Fermi-level and conductance band, respectively, and $\chi$ is the electron affinity of ML WS$_2$. Based on the potential difference $\Delta V_{CPD}$ obtained under different polarization states, corresponding Fermi-level shift can be expressed as $\Delta E_F = e\Delta V_{CPD}$, where $e$ is the elementary charge. Consequently, the ratio of carrier densities in ML WS$_2$ under opposite polarization states is given by

$$\frac{n_{up}}{n_{down}} = exp\left(\frac{\Delta E_F}{K_B T}\right) = exp(\frac{e\Delta V_{CPD}}{K_B T})$$

Where, $n_{up}$ and $n_{down}$ represent the carrier densities of ML WS$_2$ on P$_{up}$ and P$_{down}$ states, respectively. Therefore, the derived carrier density ratio is approximately 10, indicating that the carrier concentration in ML WS$_2$ can be regulated by an order of magnitude through switching of the ferroelectric polarization.

## Section 3: Mass action law

At finite doping densities, negatively or positively charged trions form by the combination of a neutral exciton with an extra electron or hole, expressed as $X^0 + e^- \rightarrow X^-$ or $X^0 + h^+ \rightarrow X^+$. The energy difference between the exciton and trion states is predicted to depend linearly on the Fermi energy and is defined as follows[8]

$$E_{X^0} - E_{X^-} = E_{X^-}^b + E_F$$

where $E_{X^-}^b$ is the trion binding energy, and $E_{X^0} - E_{X^-}$ represents the minimum energy required to remove one electron from the negative trion. $E_{X^0}$ and $E_{X^-}$ denote the energy levels of neutral excitons and trions, respectively. Hence, the trion density is directly proportional to the density of unbound carriers at the Fermi level. The unbound electron density can be theoretically estimated using the mass-action law[8-10],

$$\frac{n_{X^0} n_e}{n_{X^-}} = A k_B T \exp\left(-\frac{E_{X^-}^b}{k_B T}\right)$$

Here, $n_{X^0}$, $n_{X^-}$ and $n_e$ denote the densities of neutral excitons, negative trions, and unbound electrons, respectively. $T$ is the temperature (300 K), and $k_B$ is the Boltzmann constant. $E_{X^-}^b$ represents the binding energy of trion $X^-$. The parameter $A = \frac{4 M_{X^0} m_e}{\pi \hbar^2 M_{X^-}}$, where $M_{X^0} = m_{el} + m_{ho}$ and $M_{X^-} = 2m_{el} + m_{ho}$ are the effective masses of the neutral exciton $X^0$ and trion $X^-$, respectively. Here, $\hbar$ is the reduced Plank constant, $m_{el} = 0.31\, m_e$ and $m_{ho} = 0.42\, m_e$ are the electron and hole effective masses, with $m_e = 9.109 \times 10^{-31}\, kg$ being the electron rest mass. Likewise, for the positive trion $X^+$, the effective mass is $M_{X^+} = m_{el} + 2m_{ho}$. The trion/exciton density ratio is directly evaluated from the area ratio of the trion and exciton peaks in the PL deconvolution spectra (Fig. S13)[8,10], and the trion binding energies are also estimated from the PL deconvolution. The calculated unbound carrier densities are approximately $1.313 \times 10^{12}\, cm^{-2}$ for the $P_{up}$ state and $5.231 \times 10^{12}\, cm^{-2}$ for the $P_{down}$ state.

**Table S1 | Calculated unbound carrier densities.** Fitting parameters for PL spectral deconvolution of ML WS$_2$ on AlScN $P_{up}$ and $P_{down}$ states.

|  | Peak center (eV) | $E_{X^-}^b$ (meV) | $\frac{A_{trion}}{A_{exciton}}$ | Unbound $n_e$ (cm$^{-2}$) |
|---|---|---|---|---|
| P$_{up\_exciton}$ | 1.933 | 41 | 0.586 | $1.313 \times 10^{12}$ |
| P$_{up\_trion}$ | 1.892 | | | |
| P$_{down\_exciton}$ | 1.912 | 54 | 3.865 | $5.231 \times 10^{12}$ |
| P$_{down\_trion}$ | 1.858 | | | |

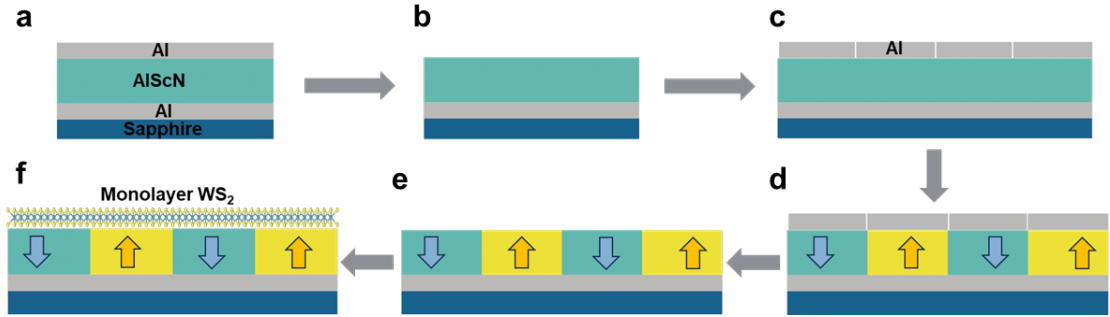

**Fig. S1 | Schematic diagram of the ML WS$_2$/patterned AlScN device fabrication process**. **a,** Initial sample structure, Al (50 nm)/AlScN (45 nm)/Al (50 nm)/Sapphire. **b,** Structure of AlScN/Al/Sapphire after etching the top Al layer. **c,** Deposition of separate Al electrodes on AlScN/Al/Sapphire for electrical poling. **d,** Application of voltages to pole AlScN into distinct polarization states. upward and downward arrows indicate AlScN poling up (P$_{up}$) and poling down (P$_{down}$), respectively. **e,** AlScN pattern after poling and subsequent etching of Al electrodes. **f,** Final device structure after transferring ML WS$_2$ onto the patterned AlScN with regions of opposite polarization.

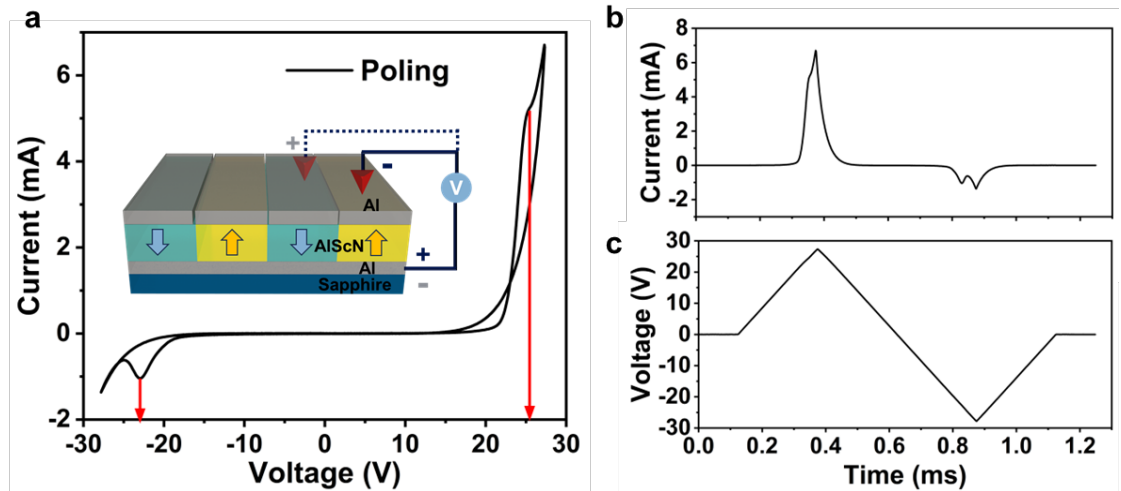

**Fig. S2 | Electrical poling of the AlScN pattern**. **a,** Current-Voltage (I-V) curve of ferroelectric AlScN during poling. Red arrows represent the coercive voltages corresponding to polarization reversal at $V_c$ = +26 V and -23 V. Inset shows a schematic of the electrical poling setup. **b,** Current-Time (I-T) curve of AlScN under applied triangular voltage pulse. **c,** Applied Voltage-Time (V-T) curve for poling. Frequency: 1 kHz.

Measurements were conducted with a standard bipolar voltage pulse with linear ramp-up and ramp-down sequences, configured as a 1 kHz triangular pulse. To achieve the polarized-up state ($P_{up}$), only positive triangular pulses were applied, conversely, negative triangular pulses were applied to obtain the polarized-down state ($P_{down}$).

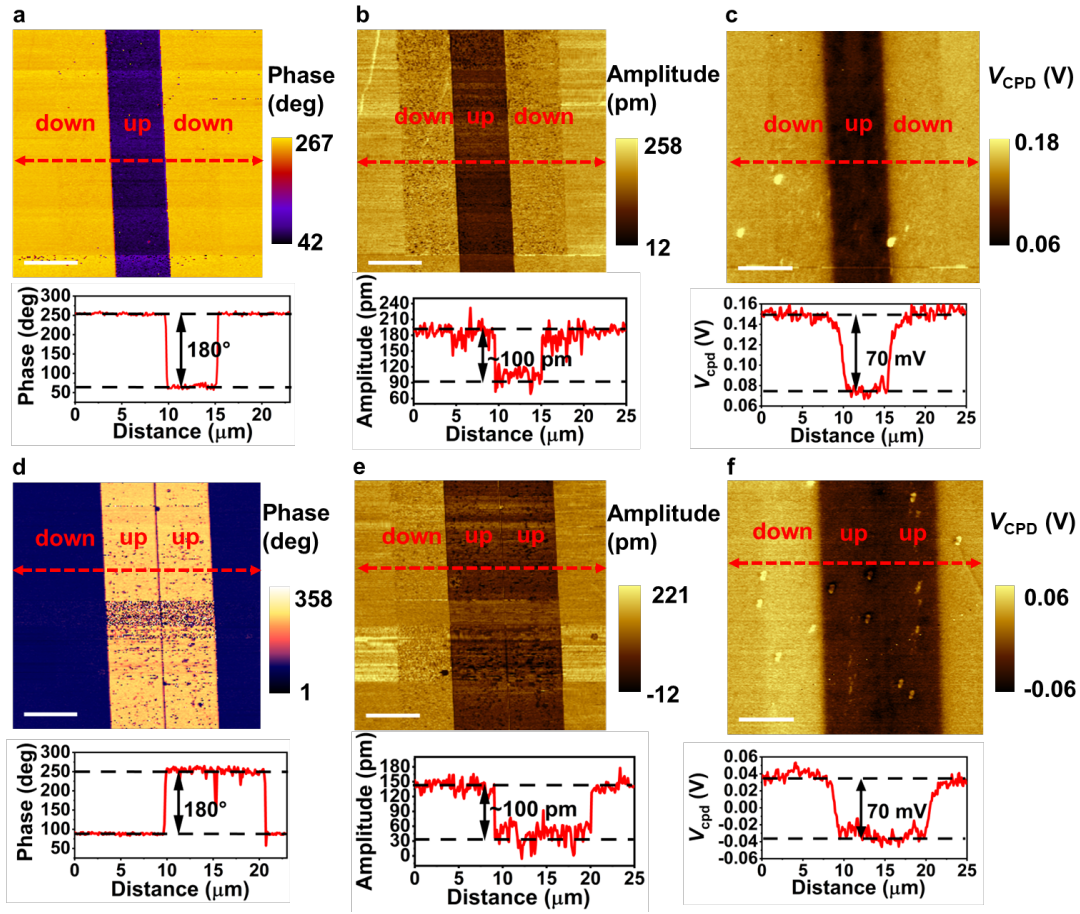

**Fig. S3 | PFM and SKPM mappings of two additional patterned AlScN**. **a-c** Phase, amplitude, and $V_{CPD}$ images of AlScN on $P_{down}$, $P_{up}$, and $P_{down}$ states, respectively. **d-f** Phase, amplitude, and $V_{CPD}$ images of AlScN on $P_{down}$, $P_{up}$, and $P_{up}$ states, respectively. These results demonstrate the reliable and stable polarization characteristics of AlScN. Scale bars, 5 μm.

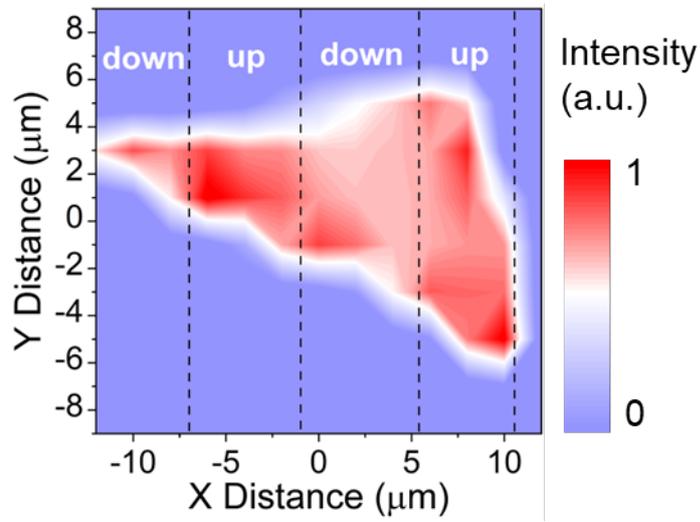

**Fig. S4 | Normalized spatial PL intensity mapping of ML WS$_2$ on patterned AlScN.** Up regions exhibit higher PL intensity compared to the down regions.

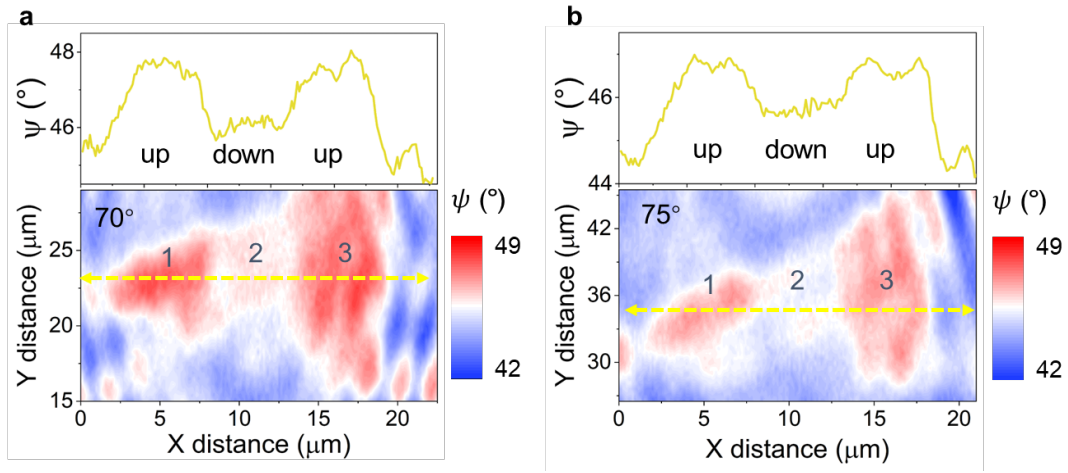

**Fig. S5 | SE mapping of $\psi$ for ML WS$_2$ on patterned AlScN at the wavelength of 636 nm (around A exciton).** (a) Angle of incidence (AOI): 70°, (b) AIO: 75°. Top profiles correspond to the yellow dashed lines in each mapping. Regions 1, 2, and 3 represent to P$_{up}$, P$_{down}$, and P$_{up}$ states, respectively.

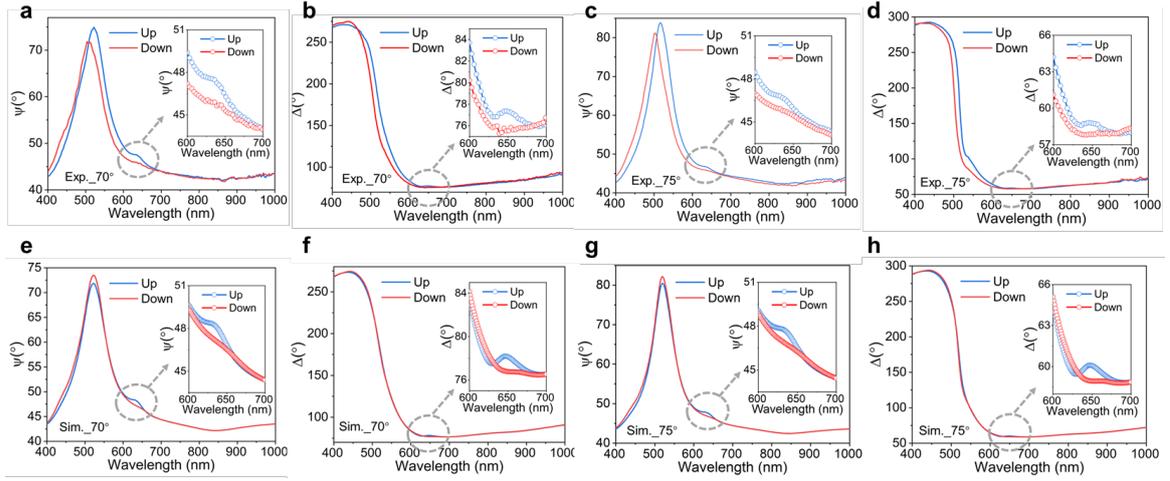

**Fig. S6 | Experimental and simulated $\psi$ and $\Delta$ spectra.** Measured $\psi$ (**c**) and $\Delta$ (**f**), alongside simulated $\psi$ (**d**) and $\Delta$ (**g**) spectra, spanning 400 to 1000 nm at incidence angles of 70° and 75°, respectively. Insets provide magnified views of the 600-700 nm range, highlighting features near the A exciton.

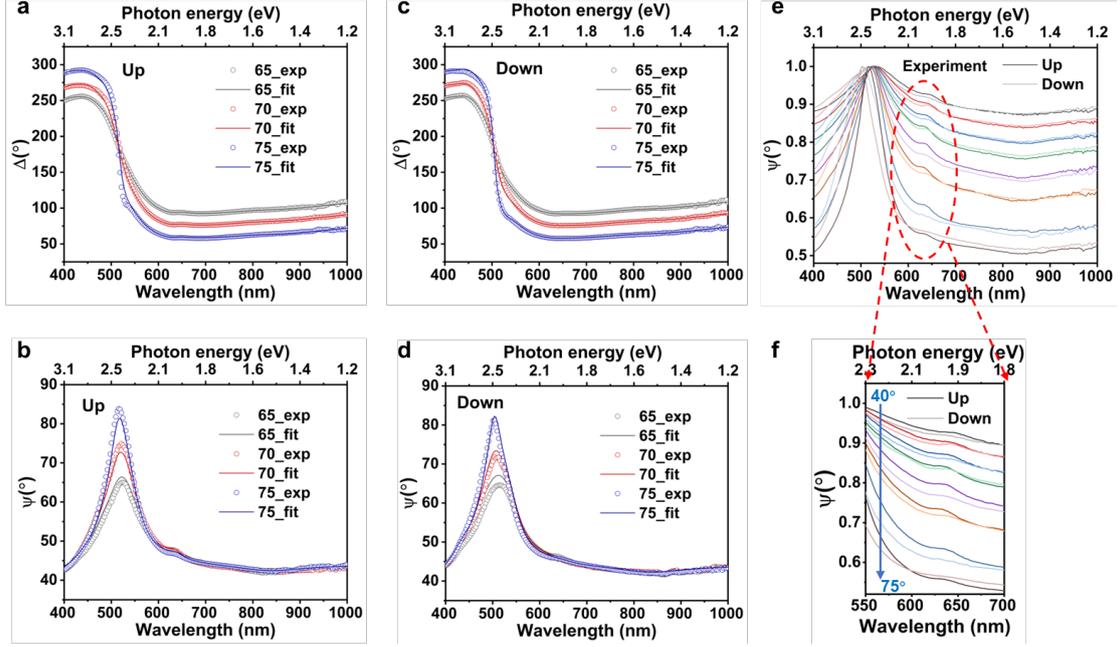

**Fig. S7 | Tauc-Lorentz fitting of $\psi$ and $\Delta$ spectra for ML WS$_2$/AlScN. a-d.** Measured (solid lines) and fitted (dashed lines) $\psi$ and $\Delta$ spectra at AOI of 65°, 70°, and 75° for P$_{up}$ state (**a,b**) and P$_{down}$ state (**c,d**). The fitting employed a multi Tauc-Lorentz oscillator model to extract the complex refractive index of ML WS$_2$ by minimizing the root mean-squared error (RMSE), defined as $RMSE = \sqrt{\frac{1}{15p-q}\sum_i \left(\psi_i^{Mod} - \psi_i^{Exp}\right)^2 + \left(\Delta_i^{Mod} - \Delta_i^{Exp}\right)^2}$, where $p$ is the number of measured wavelengths, $q$ is the number of fit parameters, and "Mod" and "Exp" denote the model and experimental values, respectively. An RMSE of ~5 was obtained, indicating excellent agreement between experiment and model. **e.** Normalized $\psi$ spectra measured at AOI from 40° to 75° in 5° step for P$_{up}$ state (solid lines) and P$_{down}$ (transparent lines) states. **f.** Zoom-in of panel (**e**) between 550 and 700 nm, highlighting features near the A exciton.

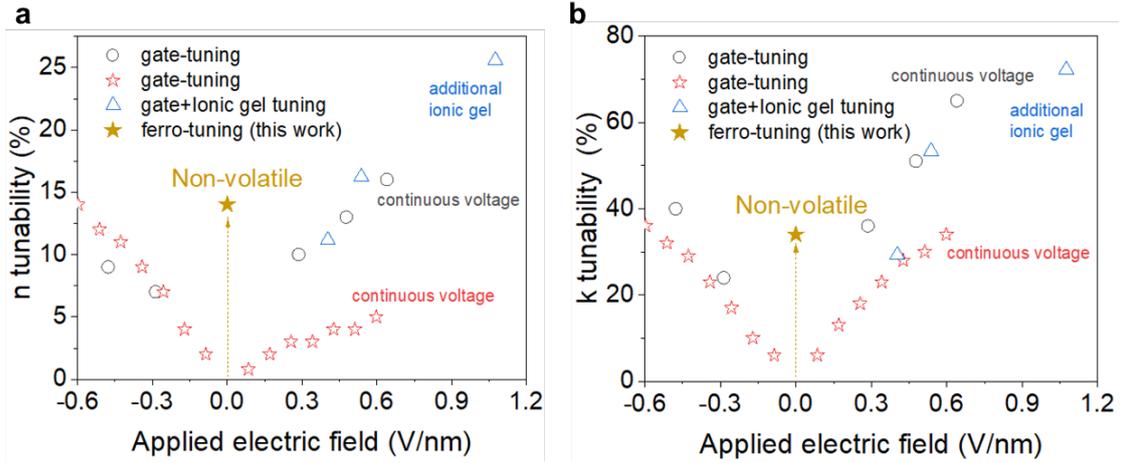

**Fig. S8 | Comparison of tunability in the optical constants of ML WS$_2$.** Tunability in n (**a**) and k (**b**) of ML WS$_2$ achieved by different tuning methods. Values were extracted from the literature[5,11,12]. For gate-tuning method, tunability is defined as $\frac{n(V)-n(0)}{n(0)}$ and $\frac{k(V)-k(0)}{k(0)}$, where $n(V)$ and $k(V)$ are the refractive indices under applied voltages, and $n(0)$ and $k(0)$ are the initial refractive indices. For our work, tunability is defined as $\frac{n(\text{up})-n(\text{down})}{n(\text{ave})}$ and $\frac{k(\text{up})-k(\text{down})}{k(\text{ave})}$, where $n(\text{ave}) = \frac{n(\text{up})+n(\text{down})}{2}$ and $k(\text{ave}) = \frac{k(\text{up})+k(\text{down})}{2}$ (the average value for P$_{up}$ and P$_{down}$ states).

**Table S2 | Comparison of tunable optical dispersion and PL in 2D system.**

| Structure | Tuning mechanism | Continuous voltage | PL tunability | n,k tunability | Non-volatile? |
|---|---|---|---|---|---|
| WS$_2$/SiO$_2$ [5] | gate-tuning | yes | yes | yes | no |
| WS$_2$/SiO$_2$ [11] | gate-tuning | yes | yes | yes | no |
| WS$_2$/ionic gel/Al$_2$O$_3$ [12] | gate+ionic gel tuning | yes | yes | yes | no |
| MoS$_2$/Si$_2$N$_4$ [13] | gate-tuning | yes | yes | yes | no |
| MoS$_2$/Al$_2$O$_3$ [14] | dopant-tuning | no | yes | yes | no |
| MoTe$_2$/P(VDF-TrFE) [15] | ferro-tuning | no | yes | no | yes |
| WSe$_2$/AlScN [16] | ferro-tuning | no | yes | no | yes |
| WS$_2$/PZT [17] | ferro-tuning | no | yes | no | yes |
| WSe$_2$/BFO [18] | ferro-tuning | no | yes | no | yes |
| MoSe$_2$(WSe$_2$)/LiNbO$_3$ [19] | ferro-tuning | no | yes | no | yes |
| MoTe$_2$/P(VDF-TrFE) [20] | ferro-tuning | no | yes | no | yes |
| WS$_2$/PZT [21] | ferro-tuning | no | yes | no | yes |
| MoSe$_2$/CuInP$_2$S$_6$ [22] | ferro-tuning | no | yes | no | yes |
| WS$_2$/AlScN (this work) | ferro-tuning | no | yes | yes | yes |

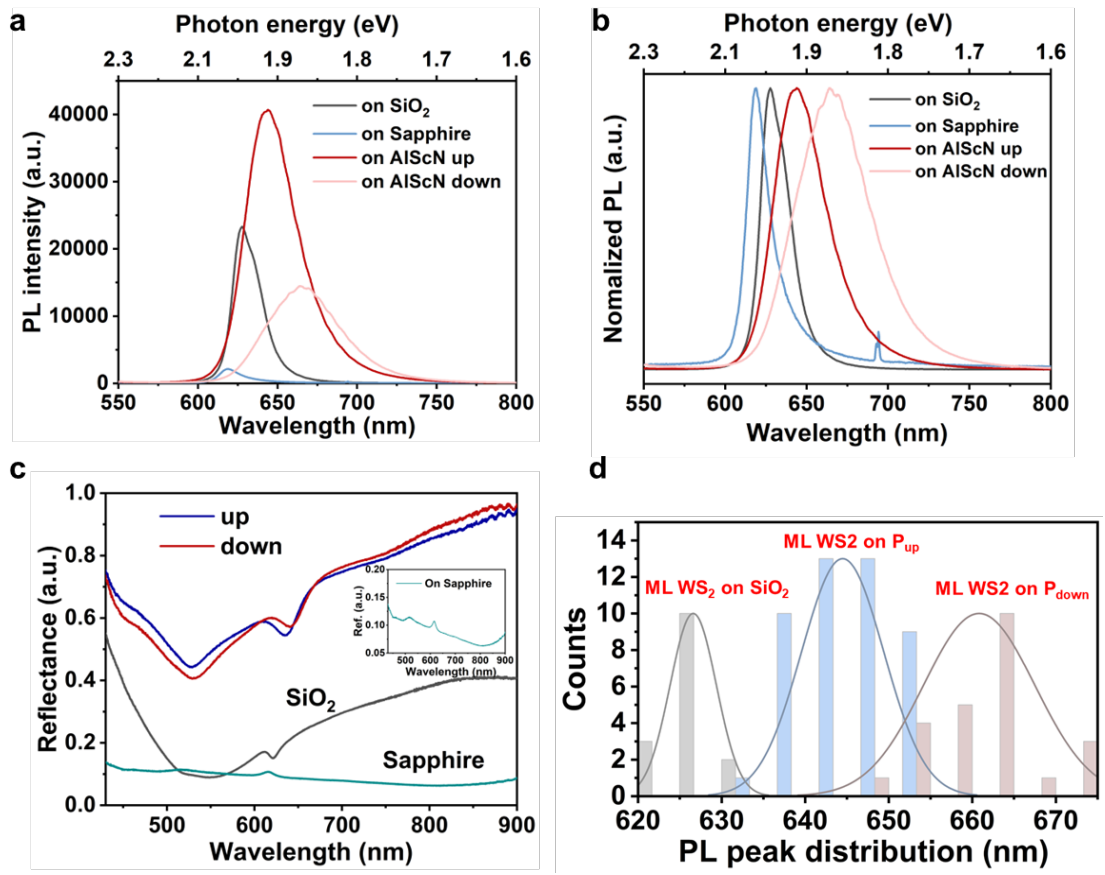

**Fig. S9 | PL and reflectance measurements of ML WS$_2$ on different substrates. a,** PL spectra acquired from different substrates. **b,** Normalized PL spectra corresponding to (**a**). **c.** Reflectance spectra acquired from different substrates. Inset: reflectance spectrum of ML WS$_2$ on sapphire. **d,** Gaussian-fitted distribution of PL peak positions for ML WS$_2$ on SiO$_2$, AlScN P$_{up}$ and P$_{down}$ (extracted from PL mapping).

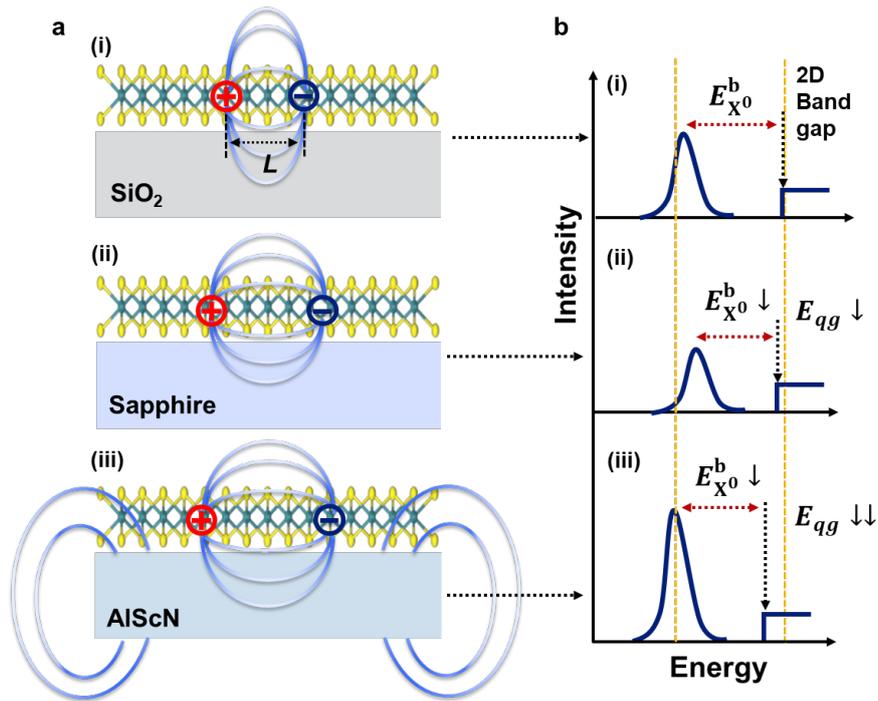

**Fig. S10 | Schematic illustration of dielectric screening effect. a.** Real-space schematic of excitons in ML WS$_2$ supported on different substrates: (i) SiO$_2$, (ii) sapphire, and (iii) AlScN. **b.** Illustration of the impact of the (i) SiO$_2$, (ii) sapphire, and (iii) AlScN substrates on the exciton binding energy ($E_{X^0}^b$) and the quasiparticle electronic bandgap ($E_{qg}$) in ML WS$_2$.

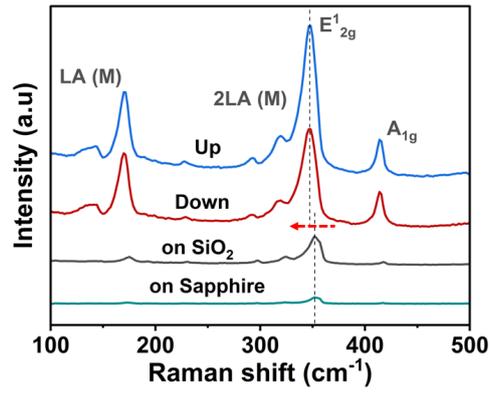

**Fig. S11 | Raman spectra of ML WS$_2$ on different substrates.** The red arrow denotes that the E$_{2g}$ peak shows a redshift comparing AlScN to SiO$_2$ and sapphire.

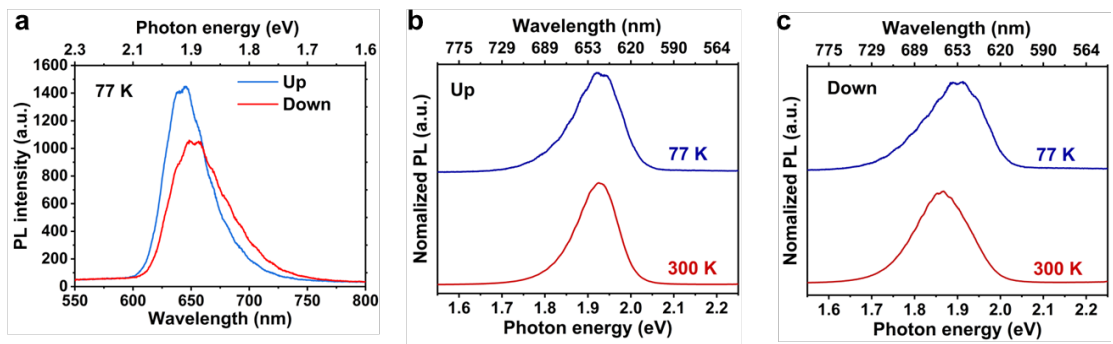

**Fig. S12 | Low-temperature PL of ML WS$_2$ on AlScN P$_{up}$ and P$_{down}$. a.** PL spectra measured at 77 K, showing a redshift of peak and a reduction of intensity from up to down. **b,c.** Comparation of normalized PL spectra at low temperature (LT) and room temperature (RT) for the AlScN P$_{up}$ (**b**) and P$_{down}$ (**c**) states, respectively.

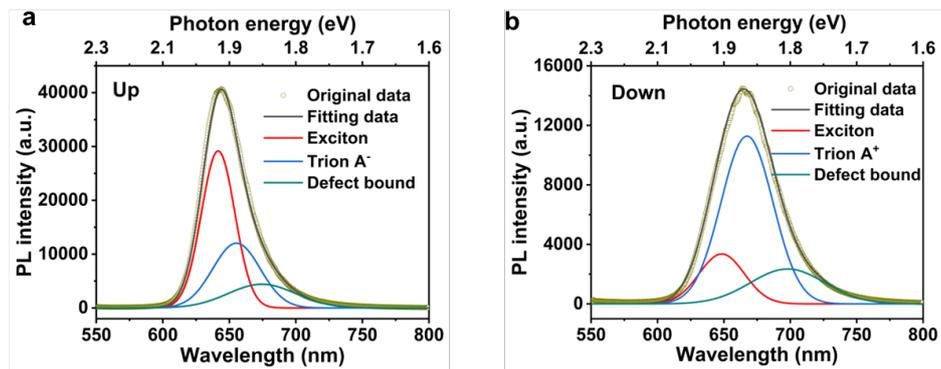

**Fig. S13 | PL spectral deconvolution of ML WS$_2$ on AlScN.** PL of ML WS$_2$ on AlScN P$_{up}$ (**a**) and P$_{down}$ (**b**) states, with deconvolution into neutral exciton and trion components.

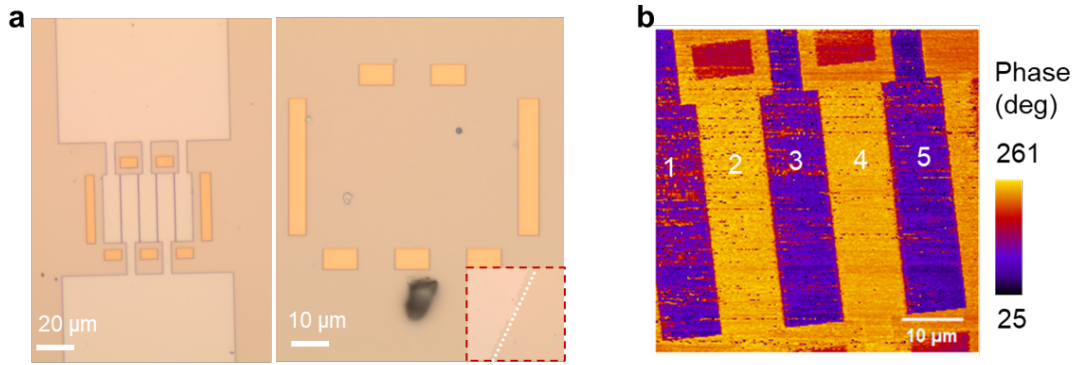

**Fig. S14 | Wet-transferred MOCVD-grown ML WS$_2$ on patterned AlScN**. **a.** Left one: OM image of AlScN pattern with five electrodes. Right one: OM image of wet-transferred ML WS$_2$ on patterned AlScN (after poling and etching the top electrodes). Inset: the boundary of AlScN and ML WS$_2$. **b.** PFM image of patterned AlScN with a 180° phase contrast between oppositely polarized domains. Labels 1, 3, and 5 correspond to P$_{up}$ states, and Labels 2 and 4 correspond to P$_{down}$ states.

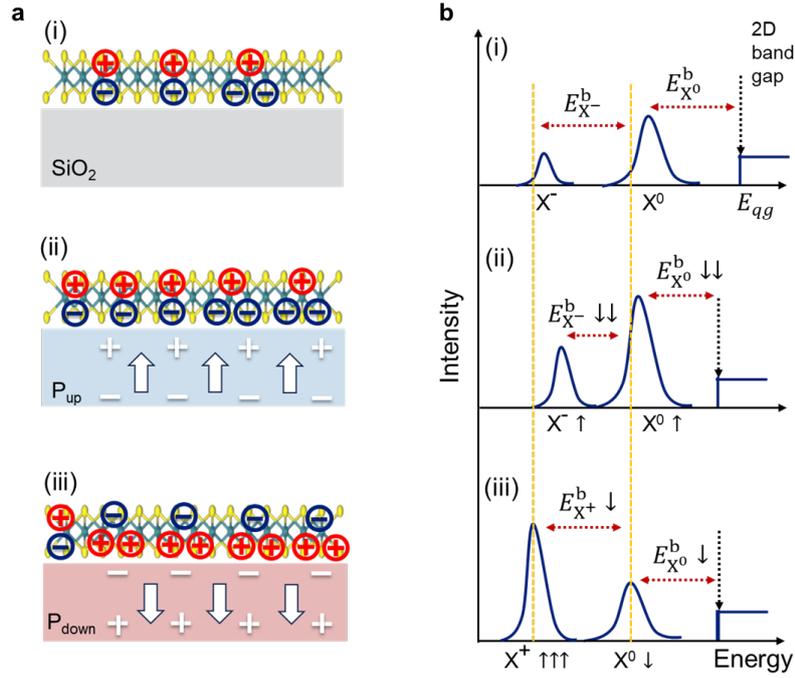

**Fig. S15 | Schematic of neutral excitons and trions in ML WS$_2$. a.** Schematic of the distribution of neutral excitons and trions in ML WS$_2$ on different substrates: (i) SiO$_2$ (no additional doping), (ii) AlScN P$_{up}$ state (electrons doping), and (iii) AlScN P$_{down}$ state (holes doping). **b.** Illustration of the impact of asymmetric screening effect on the binding energies and densities of neutral excitons and trions in ML WS$_2$, under (i) SiO$_2$, (ii) AlScN P$_{up}$, and (iii) AlScN P$_{down}$. X$^0$, X$^-$, and X$^+$ denote neutral excitons, negative trions, and positive trions. $E_{qg}$, $E_{X^0}^b$, $E_{X^-}^b$, and $E_{X^+}^b$ represent the quasiparticle bandgap, the binding energy of neutral excitons, negative trions, and positive trions, respectively.

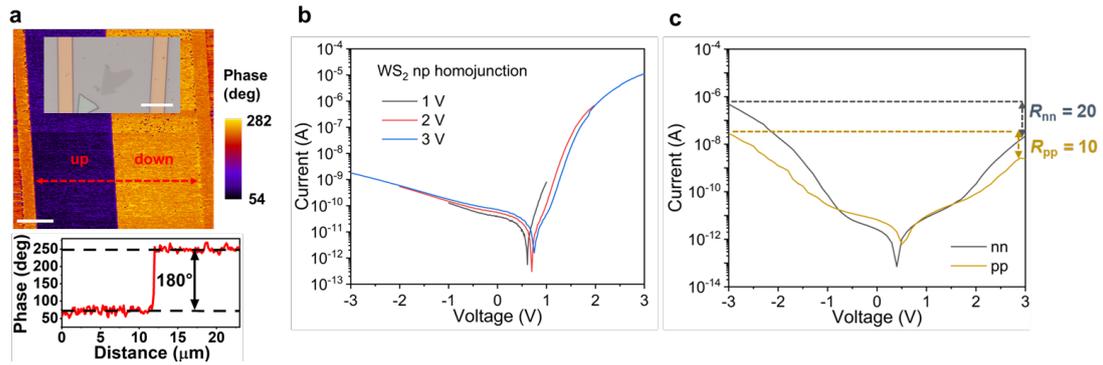

**Fig. S16 | Electrical measurements of ML WS$_2$ on AlScN. a.** PFM image and corresponding phase profile along the red dashed line, indicating a 180° phase contrast between oppositely polarized domains, scale bars, 5 µm. Inset is the optical micrograph the ML WS$_2$ flake transferred onto P$_{up}$ and P$_{down}$ regions before the electrode deposition, scale bar, 10 µm. **b.** Semi-logarithmic I-V curve of the ML WS$_2$ n-p homojunction under different bias voltages. **c.** I-V curves of the nn and pp configurations plotted on a semi-log scale. The rectification ratios at $V = \pm 3$ V are calculated as $R_{nn} = 20$ and $R_{pp} = 10$, respectively.

**Table S3 | Complex refractive index of ML WS$_2$ on AlScN under different polarization states (P$_{up}$ and P$_{down}$)**

| Wavelength (nm) | n$_{up}$ | k$_{up}$ | n$_{down}$ | k$_{down}$ |
|---|---|---|---|---|
| 400 | 4.434 | 1.575 | 3.728 | 1.861 |
| 401.5 | 4.378 | 1.624 | 3.684 | 1.929 |
| 403 | 4.33 | 1.674 | 3.648 | 1.996 |
| 404.5 | 4.289 | 1.722 | 3.62 | 2.062 |
| 406 | 4.255 | 1.77 | 3.6 | 2.127 |
| 407.5 | 4.228 | 1.817 | 3.587 | 2.19 |
| 409 | 4.206 | 1.862 | 3.58 | 2.252 |
| 410.5 | 4.191 | 1.906 | 3.58 | 2.312 |
| 412 | 4.18 | 1.949 | 3.586 | 2.37 |
| 413.5 | 4.175 | 1.989 | 3.597 | 2.426 |
| 415 | 4.175 | 2.028 | 3.614 | 2.479 |
| 416.5 | 4.179 | 2.064 | 3.635 | 2.529 |
| 418 | 4.187 | 2.098 | 3.662 | 2.576 |
| 419.5 | 4.199 | 2.129 | 3.693 | 2.62 |
| 421 | 4.215 | 2.157 | 3.728 | 2.661 |
| 422.5 | 4.234 | 2.183 | 3.768 | 2.697 |
| 424 | 4.256 | 2.205 | 3.811 | 2.731 |
| 425.5 | 4.281 | 2.223 | 3.857 | 2.76 |
| 427 | 4.308 | 2.239 | 3.907 | 2.784 |
| 428.5 | 4.337 | 2.251 | 3.959 | 2.804 |
| 430 | 4.368 | 2.259 | 4.014 | 2.819 |
| 431.5 | 4.4 | 2.264 | 4.07 | 2.829 |
| 433 | 4.432 | 2.265 | 4.128 | 2.834 |
| 434.5 | 4.466 | 2.263 | 4.187 | 2.834 |
| 436 | 4.499 | 2.256 | 4.246 | 2.828 |
| 437.5 | 4.533 | 2.246 | 4.305 | 2.816 |
| 439 | 4.565 | 2.233 | 4.363 | 2.8 |
| 440.5 | 4.597 | 2.216 | 4.419 | 2.778 |
| 442 | 4.627 | 2.196 | 4.473 | 2.751 |
| 443.5 | 4.656 | 2.174 | 4.524 | 2.719 |
| 445 | 4.682 | 2.148 | 4.572 | 2.684 |
| 446.5 | 4.707 | 2.121 | 4.616 | 2.645 |
| 448 | 4.729 | 2.091 | 4.656 | 2.603 |
| 449.5 | 4.749 | 2.06 | 4.692 | 2.559 |
| 451 | 4.766 | 2.028 | 4.724 | 2.513 |
| 452.5 | 4.78 | 1.994 | 4.751 | 2.466 |
| 454 | 4.792 | 1.961 | 4.774 | 2.419 |
| 455.5 | 4.801 | 1.927 | 4.793 | 2.372 |
| 457 | 4.808 | 1.893 | 4.809 | 2.326 |
| 458.5 | 4.812 | 1.86 | 4.821 | 2.281 |
| 460 | 4.815 | 1.827 | 4.83 | 2.238 |
| 461.5 | 4.815 | 1.795 | 4.838 | 2.196 |
| 463 | 4.814 | 1.763 | 4.843 | 2.156 |
| 464.5 | 4.812 | 1.733 | 4.847 | 2.117 |
| 466 | 4.808 | 1.704 | 4.85 | 2.08 |
| 467.5 | 4.803 | 1.676 | 4.852 | 2.044 |
| 469 | 4.797 | 1.649 | 4.854 | 2.009 |
| 470.5 | 4.79 | 1.623 | 4.855 | 1.975 |
| 472 | 4.783 | 1.598 | 4.856 | 1.941 |
| 473.5 | 4.774 | 1.573 | 4.857 | 1.908 |
| 475 | 4.765 | 1.55 | 4.857 | 1.875 |
| 476.5 | 4.756 | 1.527 | 4.855 | 1.841 |
| 478 | 4.745 | 1.504 | 4.853 | 1.807 |
| 479.5 | 4.734 | 1.483 | 4.85 | 1.773 |
| 481 | 4.722 | 1.462 | 4.844 | 1.739 |
| 482.5 | 4.709 | 1.443 | 4.837 | 1.705 |
| 484 | 4.695 | 1.424 | 4.827 | 1.671 |
| 485.5 | 4.68 | 1.407 | 4.815 | 1.639 |
| 487 | 4.664 | 1.392 | 4.801 | 1.608 |
| 488.5 | 4.647 | 1.378 | 4.784 | 1.579 |
| 490 | 4.63 | 1.366 | 4.766 | 1.551 |
| 491.5 | 4.612 | 1.357 | 4.745 | 1.527 |
| 493 | 4.595 | 1.349 | 4.723 | 1.505 |
| 494.5 | 4.578 | 1.344 | 4.7 | 1.487 |
| 496 | 4.562 | 1.342 | 4.677 | 1.472 |
| 497.5 | 4.547 | 1.341 | 4.653 | 1.461 |
| 499 | 4.533 | 1.342 | 4.63 | 1.454 |
| 500.5 | 4.521 | 1.345 | 4.608 | 1.45 |
| 502 | 4.511 | 1.35 | 4.588 | 1.449 |
| 503.5 | 4.503 | 1.356 | 4.57 | 1.451 |
| 505 | 4.498 | 1.363 | 4.555 | 1.457 |
| 506.5 | 4.495 | 1.37 | 4.542 | 1.464 |
| 508 | 4.494 | 1.378 | 4.533 | 1.474 |
| 509.5 | 4.496 | 1.385 | 4.528 | 1.485 |
| 511 | 4.5 | 1.392 | 4.526 | 1.497 |
| 512.5 | 4.507 | 1.398 | 4.529 | 1.509 |
| 514 | 4.515 | 1.402 | 4.535 | 1.52 |
| 515.5 | 4.526 | 1.406 | 4.546 | 1.53 |
| 517 | 4.538 | 1.408 | 4.56 | 1.539 |
| 518.5 | 4.551 | 1.408 | 4.577 | 1.546 |
| 520 | 4.566 | 1.407 | 4.597 | 1.549 |
| 521.5 | 4.582 | 1.403 | 4.62 | 1.55 |
| 523 | 4.598 | 1.396 | 4.645 | 1.546 |
| 524.5 | 4.614 | 1.388 | 4.671 | 1.539 |
| 526 | 4.631 | 1.377 | 4.698 | 1.527 |
| 527.5 | 4.646 | 1.364 | 4.725 | 1.512 |
| 529 | 4.662 | 1.348 | 4.75 | 1.492 |
| 530.5 | 4.676 | 1.331 | 4.775 | 1.469 |
| 532 | 4.689 | 1.313 | 4.797 | 1.442 |
| 532.1 | 4.69 | 1.311 | 4.799 | 1.44 |
| 533.5 | 4.7 | 1.292 | 4.818 | 1.413 |
| 535 | 4.711 | 1.271 | 4.835 | 1.38 |
| 536.5 | 4.72 | 1.248 | 4.85 | 1.345 |
| 538 | 4.726 | 1.224 | 4.861 | 1.309 |
| 539.5 | 4.732 | 1.199 | 4.868 | 1.272 |
| 541 | 4.735 | 1.174 | 4.873 | 1.234 |
| 542.5 | 4.737 | 1.148 | 4.874 | 1.196 |
| 544 | 4.736 | 1.122 | 4.872 | 1.158 |
| 545.5 | 4.734 | 1.097 | 4.867 | 1.121 |
| 547 | 4.73 | 1.071 | 4.859 | 1.085 |
| 548.5 | 4.724 | 1.046 | 4.849 | 1.05 |
| 550 | 4.717 | 1.021 | 4.836 | 1.016 |
| 551.5 | 4.707 | 0.998 | 4.821 | 0.984 |
| 553 | 4.697 | 0.975 | 4.805 | 0.954 |
| 554.5 | 4.685 | 0.952 | 4.787 | 0.926 |
| 556 | 4.672 | 0.931 | 4.768 | 0.899 |
| 557.5 | 4.657 | 0.911 | 4.747 | 0.875 |
| 559 | 4.642 | 0.892 | 4.726 | 0.852 |
| 560.5 | 4.625 | 0.875 | 4.704 | 0.831 |
| 562 | 4.607 | 0.858 | 4.681 | 0.812 |
| 563.5 | 4.589 | 0.843 | 4.658 | 0.795 |
| 565 | 4.57 | 0.83 | 4.635 | 0.779 |
| 566.5 | 4.55 | 0.817 | 4.611 | 0.765 |
| 568 | 4.53 | 0.806 | 4.588 | 0.753 |
| 569.5 | 4.509 | 0.797 | 4.564 | 0.743 |
| 571 | 4.488 | 0.789 | 4.54 | 0.734 |
| 572.5 | 4.466 | 0.783 | 4.517 | 0.726 |
| 574 | 4.445 | 0.778 | 4.494 | 0.72 |
| 575.5 | 4.423 | 0.775 | 4.471 | 0.716 |
| 577 | 4.4 | 0.773 | 4.449 | 0.712 |
| 578.5 | 4.378 | 0.773 | 4.427 | 0.71 |
| 580 | 4.356 | 0.775 | 4.405 | 0.71 |
| 581.5 | 4.334 | 0.778 | 4.384 | 0.71 |
| 583 | 4.312 | 0.783 | 4.363 | 0.712 |
| 584.5 | 4.29 | 0.79 | 4.343 | 0.715 |
| 586 | 4.269 | 0.799 | 4.324 | 0.72 |
| 587.5 | 4.248 | 0.809 | 4.305 | 0.725 |
| 589 | 4.228 | 0.822 | 4.287 | 0.731 |
| 590.5 | 4.209 | 0.836 | 4.27 | 0.739 |
| 592 | 4.19 | 0.853 | 4.254 | 0.748 |
| 593.5 | 4.172 | 0.872 | 4.238 | 0.757 |
| 595 | 4.156 | 0.893 | 4.224 | 0.768 |
| 596.5 | 4.141 | 0.916 | 4.21 | 0.78 |
| 598 | 4.128 | 0.942 | 4.197 | 0.792 |
| 599.5 | 4.116 | 0.969 | 4.186 | 0.806 |
| 601 | 4.107 | 0.999 | 4.175 | 0.821 |
| 602.5 | 4.1 | 1.032 | 4.166 | 0.836 |
| 604 | 4.097 | 1.066 | 4.158 | 0.853 |
| 605.5 | 4.096 | 1.103 | 4.151 | 0.87 |
| 607 | 4.1 | 1.142 | 4.147 | 0.888 |
| 608.5 | 4.107 | 1.182 | 4.143 | 0.907 |
| 610 | 4.12 | 1.224 | 4.141 | 0.926 |
| 611.5 | 4.138 | 1.266 | 4.141 | 0.946 |
| 613 | 4.161 | 1.309 | 4.143 | 0.967 |
| 614.5 | 4.19 | 1.351 | 4.147 | 0.988 |
| 616 | 4.225 | 1.393 | 4.153 | 1.009 |
| 617.5 | 4.268 | 1.432 | 4.161 | 1.03 |
| 619 | 4.316 | 1.468 | 4.171 | 1.052 |
| 620.5 | 4.373 | 1.5 | 4.185 | 1.072 |
| 622 | 4.433 | 1.527 | 4.199 | 1.093 |
| 623.5 | 4.501 | 1.547 | 4.218 | 1.113 |
| 625 | 4.571 | 1.56 | 4.237 | 1.132 |
| 626.5 | 4.647 | 1.563 | 4.261 | 1.15 |
| 628 | 4.723 | 1.557 | 4.285 | 1.166 |
| 629.5 | 4.801 | 1.542 | 4.314 | 1.18 |
| 631 | 4.877 | 1.516 | 4.343 | 1.193 |
| 632.5 | 4.95 | 1.48 | 4.376 | 1.203 |
| 634 | 5.019 | 1.435 | 4.41 | 1.21 |
| 635.5 | 5.081 | 1.382 | 4.446 | 1.215 |
| 637 | 5.135 | 1.321 | 4.483 | 1.216 |
| 638.5 | 5.183 | 1.255 | 4.521 | 1.214 |
| 640 | 5.22 | 1.187 | 4.561 | 1.208 |
| 641.5 | 5.253 | 1.114 | 4.6 | 1.198 |
| 643 | 5.271 | 1.043 | 4.64 | 1.184 |
| 644.5 | 5.289 | 0.97 | 4.678 | 1.167 |
| 646 | 5.291 | 0.901 | 4.716 | 1.146 |
| 647.5 | 5.294 | 0.833 | 4.752 | 1.12 |
| 649 | 5.287 | 0.769 | 4.785 | 1.092 |
| 650.5 | 5.277 | 0.709 | 4.817 | 1.06 |
| 652 | 5.263 | 0.652 | 4.845 | 1.026 |
| 653.5 | 5.244 | 0.599 | 4.871 | 0.989 |
| 655 | 5.226 | 0.549 | 4.894 | 0.95 |
| 656.5 | 5.202 | 0.504 | 4.912 | 0.91 |
| 658 | 5.179 | 0.462 | 4.929 | 0.868 |
| 659.5 | 5.154 | 0.423 | 4.939 | 0.826 |
| 661 | 5.128 | 0.388 | 4.948 | 0.784 |
| 662.5 | 5.103 | 0.355 | 4.954 | 0.742 |
| 664 | 5.076 | 0.326 | 4.956 | 0.7 |
| 665.5 | 5.05 | 0.298 | 4.957 | 0.659 |
| 667 | 5.024 | 0.273 | 4.952 | 0.62 |
| 668.5 | 4.998 | 0.251 | 4.947 | 0.581 |
| 670 | 4.973 | 0.23 | 4.939 | 0.544 |
| 671.5 | 4.948 | 0.211 | 4.929 | 0.509 |
| 673 | 4.923 | 0.193 | 4.918 | 0.475 |
| 674.5 | 4.899 | 0.177 | 4.904 | 0.442 |
| 676 | 4.876 | 0.163 | 4.89 | 0.412 |
| 677.5 | 4.853 | 0.149 | 4.875 | 0.383 |
| 679 | 4.831 | 0.137 | 4.859 | 0.356 |
| 680.5 | 4.808 | 0.125 | 4.842 | 0.33 |
| 682 | 4.788 | 0.115 | 4.824 | 0.306 |
| 683.5 | 4.767 | 0.105 | 4.807 | 0.284 |
| 685 | 4.747 | 0.097 | 4.789 | 0.262 |
| 686.5 | 4.727 | 0.089 | 4.77 | 0.243 |
| 688 | 4.708 | 0.081 | 4.752 | 0.224 |
| 689.5 | 4.69 | 0.074 | 4.733 | 0.207 |
| 691 | 4.672 | 0.068 | 4.715 | 0.191 |
| 692.5 | 4.654 | 0.062 | 4.696 | 0.176 |
| 694 | 4.638 | 0.057 | 4.678 | 0.162 |
| 695.5 | 4.621 | 0.052 | 4.66 | 0.15 |
| 697 | 4.605 | 0.047 | 4.642 | 0.138 |
| 698.5 | 4.59 | 0.043 | 4.624 | 0.127 |
| 700 | 4.575 | 0.039 | 4.607 | 0.116 |
| 701.5 | 4.56 | 0.036 | 4.59 | 0.107 |
| 703 | 4.546 | 0.033 | 4.573 | 0.098 |
| 704.5 | 4.532 | 0.03 | 4.557 | 0.09 |
| 706 | 4.518 | 0.027 | 4.541 | 0.082 |
| 707.5 | 4.505 | 0.025 | 4.525 | 0.075 |
| 709 | 4.492 | 0.022 | 4.509 | 0.068 |
| 710.5 | 4.48 | 0.02 | 4.494 | 0.062 |
| 712 | 4.468 | 0.018 | 4.48 | 0.057 |
| 713.5 | 4.456 | 0.016 | 4.465 | 0.051 |
| 715 | 4.444 | 0.015 | 4.451 | 0.047 |
| 716.5 | 4.433 | 0.013 | 4.437 | 0.042 |
| 718 | 4.422 | 0.012 | 4.423 | 0.038 |
| 719.5 | 4.411 | 0.01 | 4.41 | 0.034 |
| 721 | 4.401 | 0.009 | 4.397 | 0.031 |
| 722.5 | 4.39 | 0.008 | 4.384 | 0.027 |
| 724 | 4.38 | 0.007 | 4.372 | 0.024 |
| 725.5 | 4.37 | 0.006 | 4.36 | 0.022 |
| 727 | 4.361 | 0.005 | 4.348 | 0.019 |
| 728.5 | 4.351 | 0.004 | 4.336 | 0.017 |
| 730 | 4.342 | 0.004 | 4.325 | 0.015 |
| 731.5 | 4.333 | 0.003 | 4.314 | 0.013 |
| 733 | 4.324 | 0.003 | 4.303 | 0.011 |
| 734.5 | 4.316 | 0.002 | 4.293 | 0.01 |
| 736 | 4.308 | 0.002 | 4.282 | 0.008 |
| 737.5 | 4.299 | 0.001 | 4.272 | 0.007 |
| 739 | 4.291 | 0.001 | 4.262 | 0.006 |
| 740.5 | 4.284 | 0.001 | 4.253 | 0.005 |
| 742 | 4.276 | 0.001 | 4.243 | 0.004 |
| 743.5 | 4.268 | 0 | 4.234 | 0.003 |
| 745 | 4.261 | 0 | 4.225 | 0.002 |
| 746.5 | 4.254 | 0 | 4.216 | 0.002 |
| 748 | 4.247 | 5.45E-05 | 4.208 | 0.001 |
| 749.5 | 4.24 | 1.19E-05 | 4.199 | 0.001 |
| 751 | 4.233 | 0 | 4.191 | 0.001 |
| 752.5 | 4.227 | 0 | 4.183 | 0 |
| 754 | 4.22 | 0 | 4.175 | 0 |
| 755.5 | 4.214 | 0 | 4.168 | 5.49E-05 |
| 757 | 4.208 | 0 | 4.16 | 3.62E-06 |
| 758.5 | 4.201 | 0 | 4.153 | 0 |
| 760 | 4.196 | 0 | 4.146 | 0 |
| 761.5 | 4.19 | 0 | 4.139 | 0 |
| 763 | 4.184 | 0 | 4.133 | 0 |
| 764.5 | 4.178 | 0 | 4.126 | 0 |
| 766 | 4.173 | 0 | 4.12 | 0 |
| 767.5 | 4.167 | 0 | 4.113 | 0 |
| 769 | 4.162 | 0 | 4.107 | 0 |
| 770.5 | 4.157 | 0 | 4.101 | 0 |
| 772 | 4.151 | 0 | 4.095 | 0 |
| 773.5 | 4.146 | 0 | 4.09 | 0 |
| 775 | 4.141 | 0 | 4.084 | 0 |
| 776.5 | 4.136 | 0 | 4.078 | 0 |
| 778 | 4.131 | 0 | 4.073 | 0 |
| 779.5 | 4.127 | 0 | 4.067 | 0 |
| 781 | 4.122 | 0 | 4.062 | 0 |
| 782.5 | 4.117 | 0 | 4.057 | 0 |
| 784 | 4.113 | 0 | 4.052 | 0 |
| 785.5 | 4.108 | 0 | 4.047 | 0 |
| 787 | 4.104 | 0 | 4.042 | 0 |
| 788.5 | 4.099 | 0 | 4.037 | 0 |
| 790 | 4.095 | 0 | 4.032 | 0 |
| 791.5 | 4.091 | 0 | 4.027 | 0 |
| 793 | 4.086 | 0 | 4.022 | 0 |
| 794.5 | 4.082 | 0 | 4.018 | 0 |
| 796 | 4.078 | 0 | 4.013 | 0 |
| 797.5 | 4.074 | 0 | 4.009 | 0 |
| 799 | 4.07 | 0 | 4.004 | 0 |
| 800.5 | 4.066 | 0 | 4 | 0 |
| 802 | 4.062 | 0 | 3.996 | 0 |
| 803.5 | 4.058 | 0 | 3.992 | 0 |
| 805 | 4.054 | 0 | 3.987 | 0 |
| 806.5 | 4.05 | 0 | 3.983 | 0 |
| 808 | 4.047 | 0 | 3.979 | 0 |
| 809.5 | 4.043 | 0 | 3.975 | 0 |
| 811 | 4.039 | 0 | 3.971 | 0 |
| 812.5 | 4.036 | 0 | 3.967 | 0 |
| 814 | 4.032 | 0 | 3.963 | 0 |
| 815.5 | 4.029 | 0 | 3.96 | 0 |
| 817 | 4.025 | 0 | 3.956 | 0 |
| 818.5 | 4.022 | 0 | 3.952 | 0 |
| 820 | 4.018 | 0 | 3.948 | 0 |
| 821.5 | 4.015 | 0 | 3.945 | 0 |
| 823 | 4.012 | 0 | 3.941 | 0 |
| 824.5 | 4.008 | 0 | 3.938 | 0 |
| 826 | 4.005 | 0 | 3.934 | 0 |
| 827.5 | 4.002 | 0 | 3.931 | 0 |
| 829 | 3.999 | 0 | 3.927 | 0 |
| 830.5 | 3.996 | 0 | 3.924 | 0 |
| 832 | 3.992 | 0 | 3.92 | 0 |
| 833.5 | 3.989 | 0 | 3.917 | 0 |
| 835 | 3.986 | 0 | 3.914 | 0 |
| 836.5 | 3.983 | 0 | 3.911 | 0 |
| 838 | 3.98 | 0 | 3.907 | 0 |
| 839.5 | 3.977 | 0 | 3.904 | 0 |
| 841 | 3.974 | 0 | 3.901 | 0 |
| 842.5 | 3.971 | 0 | 3.898 | 0 |
| 844 | 3.969 | 0 | 3.895 | 0 |
| 845.5 | 3.966 | 0 | 3.892 | 0 |
| 847 | 3.963 | 0 | 3.889 | 0 |
| 848.5 | 3.96 | 0 | 3.886 | 0 |
| 850 | 3.957 | 0 | 3.883 | 0 |
| 851.5 | 3.955 | 0 | 3.88 | 0 |
| 853 | 3.952 | 0 | 3.877 | 0 |
| 854.5 | 3.949 | 0 | 3.874 | 0 |
| 856 | 3.947 | 0 | 3.871 | 0 |
| 857.5 | 3.944 | 0 | 3.869 | 0 |
| 859 | 3.941 | 0 | 3.866 | 0 |
| 860.5 | 3.939 | 0 | 3.863 | 0 |
| 862 | 3.936 | 0 | 3.86 | 0 |
| 863.5 | 3.934 | 0 | 3.858 | 0 |
| 865 | 3.931 | 0 | 3.855 | 0 |
| 866.5 | 3.929 | 0 | 3.852 | 0 |
| 868 | 3.926 | 0 | 3.85 | 0 |
| 869.5 | 3.924 | 0 | 3.847 | 0 |
| 871 | 3.921 | 0 | 3.845 | 0 |
| 872.5 | 3.919 | 0 | 3.842 | 0 |
| 874 | 3.917 | 0 | 3.84 | 0 |
| 875.5 | 3.914 | 0 | 3.837 | 0 |
| 877 | 3.912 | 0 | 3.835 | 0 |
| 878.5 | 3.91 | 0 | 3.832 | 0 |
| 880 | 3.907 | 0 | 3.83 | 0 |
| 881.5 | 3.905 | 0 | 3.827 | 0 |
| 883 | 3.903 | 0 | 3.825 | 0 |
| 884.5 | 3.901 | 0 | 3.823 | 0 |
| 886 | 3.898 | 0 | 3.82 | 0 |
| 887.5 | 3.896 | 0 | 3.818 | 0 |
| 889 | 3.894 | 0 | 3.816 | 0 |
| 890.5 | 3.892 | 0 | 3.813 | 0 |
| 892 | 3.89 | 0 | 3.811 | 0 |
| 893.5 | 3.888 | 0 | 3.809 | 0 |
| 895 | 3.885 | 0 | 3.807 | 0 |
| 896.5 | 3.883 | 0 | 3.804 | 0 |
| 898 | 3.881 | 0 | 3.802 | 0 |
| 899.5 | 3.879 | 0 | 3.8 | 0 |
| 901 | 3.877 | 0 | 3.798 | 0 |
| 902.5 | 3.875 | 0 | 3.796 | 0 |
| 904 | 3.873 | 0 | 3.794 | 0 |
| 905.5 | 3.871 | 0 | 3.791 | 0 |
| 907 | 3.869 | 0 | 3.789 | 0 |
| 908.5 | 3.867 | 0 | 3.787 | 0 |
| 910 | 3.865 | 0 | 3.785 | 0 |
| 911.5 | 3.863 | 0 | 3.783 | 0 |
| 913 | 3.862 | 0 | 3.781 | 0 |
| 914.5 | 3.86 | 0 | 3.779 | 0 |
| 916 | 3.858 | 0 | 3.777 | 0 |
| 917.5 | 3.856 | 0 | 3.775 | 0 |
| 919 | 3.854 | 0 | 3.773 | 0 |
| 920.5 | 3.852 | 0 | 3.771 | 0 |
| 922 | 3.85 | 0 | 3.77 | 0 |
| 923.5 | 3.849 | 0 | 3.768 | 0 |
| 925 | 3.847 | 0 | 3.766 | 0 |
| 926.5 | 3.845 | 0 | 3.764 | 0 |
| 928 | 3.843 | 0 | 3.762 | 0 |
| 929.5 | 3.841 | 0 | 3.76 | 0 |
| 931 | 3.84 | 0 | 3.758 | 0 |
| 932.5 | 3.838 | 0 | 3.757 | 0 |
| 934 | 3.836 | 0 | 3.755 | 0 |
| 935.5 | 3.835 | 0 | 3.753 | 0 |
| 937 | 3.833 | 0 | 3.751 | 0 |
| 938.5 | 3.831 | 0 | 3.749 | 0 |
| 940 | 3.83 | 0 | 3.748 | 0 |
| 941.5 | 3.828 | 0 | 3.746 | 0 |
| 943 | 3.826 | 0 | 3.744 | 0 |
| 944.5 | 3.825 | 0 | 3.743 | 0 |
| 946 | 3.823 | 0 | 3.741 | 0 |
| 947.5 | 3.821 | 0 | 3.739 | 0 |
| 949 | 3.82 | 0 | 3.737 | 0 |
| 950.5 | 3.818 | 0 | 3.736 | 0 |
| 952 | 3.817 | 0 | 3.734 | 0 |
| 953.5 | 3.815 | 0 | 3.733 | 0 |
| 955 | 3.814 | 0 | 3.731 | 0 |
| 956.5 | 3.812 | 0 | 3.729 | 0 |
| 958 | 3.811 | 0 | 3.728 | 0 |
| 959.5 | 3.809 | 0 | 3.726 | 0 |
| 961 | 3.808 | 0 | 3.725 | 0 |
| 962.5 | 3.806 | 0 | 3.723 | 0 |
| 964 | 3.805 | 0 | 3.721 | 0 |
| 965.5 | 3.803 | 0 | 3.72 | 0 |
| 967 | 3.802 | 0 | 3.718 | 0 |
| 968.5 | 3.8 | 0 | 3.717 | 0 |
| 970 | 3.799 | 0 | 3.715 | 0 |
| 971.5 | 3.797 | 0 | 3.714 | 0 |
| 973 | 3.796 | 0 | 3.712 | 0 |
| 974.5 | 3.794 | 0 | 3.711 | 0 |
| 976 | 3.793 | 0 | 3.709 | 0 |
| 977.5 | 3.792 | 0 | 3.708 | 0 |
| 979 | 3.79 | 0 | 3.706 | 0 |
| 980.5 | 3.789 | 0 | 3.705 | 0 |
| 982 | 3.787 | 0 | 3.704 | 0 |
| 983.5 | 3.786 | 0 | 3.702 | 0 |
| 985 | 3.785 | 0 | 3.701 | 0 |
| 986.5 | 3.783 | 0 | 3.699 | 0 |
| 988 | 3.782 | 0 | 3.698 | 0 |
| 989.5 | 3.781 | 0 | 3.697 | 0 |
| 991 | 3.779 | 0 | 3.695 | 0 |
| 992.5 | 3.778 | 0 | 3.694 | 0 |
| 994 | 3.777 | 0 | 3.692 | 0 |
| 995.5 | 3.776 | 0 | 3.691 | 0 |
| 997 | 3.774 | 0 | 3.69 | 0 |
| 998.5 | 3.773 | 0 | 3.688 | 0 |
| 1000 | 3.772 | 0 | 3.687 | 0 |

**References**

1   Kylänpää, I. & Komsa, H.-P. Binding energies of exciton complexes in transition metal dichalcogenide monolayers and effect of dielectric environment. *Phys. Rev. B* **92**, 205418 (2015).

2   Druppel, M., Deilmann, T., Kruger, P. & Rohlfing, M. Diversity of trion states and substrate effects in the optical properties of an $MoS_2$ monolayer. *Nat. Commun.* **8**, 2117 (2017).

3   Lin, Y. *et al.* Dielectric screening of excitons and trions in single-layer $MoS_2$. *Nano Lett.* **14**, 5569-5576 (2014).

4   Dhiren K. Pradhan, D. C. M., Gwangwoo Kim, Yunfei He, Pariasadat Musavigharavi, Kwan-Ho Kim, Nishant Sharma, Zirun Han, Xingyu Du, Venkata S. Puli, Eric A. Stach, W. Joshua Kennedy, Nicholas R. Glavin, Roy H. Olsson III & Deep Jariwala A scalable ferroelectric non-volatile memory operating at 600 °C. *Nat. Electron.* **7**, 348–355 (2024).

5   Yu, Y. *et al.* Giant Gating Tunability of Optical Refractive Index in Transition Metal Dichalcogenide Monolayers. *Nano Lett.* **17**, 3613-3618 (2017).

6   Ming-Yang Li, Y. S., Chia-Chin Cheng, Li-Syuan Lu, Yung-Chang Lin, Hao-Lin Tang, Meng-Lin Tsai, Chih-Wei Chu, Kung-Hwa Wei. Epitaxial growth of a monolayer $WSe_2$-$MoS_2$ lateral p-n junction with an atomically sharp interface. *Science* **349**, 524-528 (2015).

7   Polak, L. & Wijngaarden, R. J. in *Kelvin Probe Force Microscopy: From Single*

*Charge Detection to Device Characterization* (eds Sascha Sadewasser & Thilo Glatzel) 227-247 (Springer International Publishing, 2018).

8   Mak, K. F. *et al.* Tightly bound trions in monolayer $MoS_2$. *Nat. Mater.* **12**, 207-211 (2013).

9   Harats, M. G., Kirchhof, J. N., Qiao, M., Greben, K. & Bolotin, K. I. Dynamics and efficient conversion of excitons to trions in non-uniformly strained monolayer $WS_2$. *Nat. Photonics* **14**, 324-329 (2020).

10  Rahul Kesarwani, K. B. S., Teng-De Huang, Yu-Fan Chiang, Nai-Chang Yeh, Yann-Wen Lan, Ting-Hua Lu. Control of trion-to-exciton conversion in monolayer $WS_2$ by orbital angular momentum of light. *Sci. Adv.* **8**, eabm0100 (2022).

11  Lynch, J. *et al.* Full $2\pi$ phase modulation using exciton-polaritons in a two-dimensional superlattice. *Device* **3**, 100639 (2025).

12  Wang, Z. *et al.* Greatly Enhanced Resonant Exciton-Trion Conversion in Electrically Modulated Atomically Thin $WS_2$ at Room Temperature. *Adv. Mater.* **35**, e2302248 (2023).

13  Kravets, V. G. *et al.* Measurements of electrically tunable refractive index of $MoS_2$ monolayer and its usage in optical modulators. *npj 2D Mater. Appl.* **3**, 36 (2019).

14  Stevenson, P. R. *et al.* Reversibly Tailoring Optical Constants of Monolayer Transition Metal Dichalcogenide $MoS_2$ Films: Impact of Dopant-Induced Screening from Chemical Adsorbates and Mild Film Degradation. *ACS Photonics* **8**, 1705-1717 (2021).

15  Wu, G. *et al.* Programmable transition metal dichalcogenide homojunctions controlled by nonvolatile ferroelectric domains. *Nat. Electron.* **3**, 43-50 (2020).

16  Singh, S. *et al.* Nonvolatile Control of Valley Polarized Emission in 2D $WSe_2$-AlScN Heterostructures. *ACS Nano* **18**, 17958-17968 (2024).

17  Li, C. H., McCreary, K. M. & Jonker, B. T. Spatial Control of Photoluminescence at Room Temperature by Ferroelectric Domains in Monolayer $WS_2$/PZT Hybrid Structures. *ACS Omega* **1**, 1075-1080 (2016).

18  Chen, J. W. *et al.* A gate-free monolayer $WSe_2$ pn diode. *Nat. Commun.* **9**, 3143 (2018).

19  Wen, B. *et al.* Ferroelectric-Driven Exciton and Trion Modulation in Monolayer Molybdenum and Tungsten Diselenides. *ACS Nano* **13**, 5335-5343 (2019).

20  Wu, G. *et al.* Ferroelectric-defined reconfigurable homojunctions for in-memory sensing and computing. *Nat. Mater.* **22**, 1499-1506 (2023).

21  Qiu, X. *et al.* In-Plane Polarization-Triggered $WS_2$-Ferroelectric Heterostructured Synaptic Devices. *ACS Appl. Mater. Interfaces* **17**, 7027-7035 (2025).

22  Mao, X. *et al.* Nonvolatile Electric Control of Exciton Complexes in Monolayer $MoSe_2$ with Two-Dimensional Ferroelectric $CuInP_2S_6$. *ACS Appl. Mater. Interfaces* **13**, 24250-24257 (2021).